\documentclass[11pt]{article}      
\usepackage[margin=1in]{geometry}  
\usepackage{microtype}
\usepackage[export]{adjustbox}
\usepackage{xspace}
\usepackage{amsmath}
\usepackage{amsthm}
\usepackage{wrapfig}
\usepackage{thm-restate}
\usepackage[font=small]{caption}
\usepackage[hidelinks]{hyperref}
\usepackage[capitalise]{cleveref}
\usepackage{amsfonts}
\usepackage{amssymb}
\usepackage[textsize=tiny]{todonotes}
\usepackage[inline]{enumitem}
\usepackage{mathtools}
\usepackage{cases}
\usepackage[only,llbracket,rrbracket]{stmaryrd}

\usepackage{array}
\usepackage{multirow}
\usepackage{multicol}
\usepackage{mathrsfs} 
\newtheorem{theorem}{Theorem}[section]
\newtheorem{corollary}[theorem]{Corollary}
\newtheorem{lemma}[theorem]{Lemma}
\newtheorem{claim}[theorem]{Claim}
\newtheorem{proposition}[theorem]{Proposition}
\newtheorem{conjecture}[theorem]{Conjecture}
\theoremstyle{definition}
\newtheorem{definition}[theorem]{Definition}
\newtheorem{observation}[theorem]{Observation}

\usetikzlibrary{arrows.meta}

\usepackage{enumitem}
\setlist[itemize]{topsep=4pt,itemsep=3pt,parsep=0pt} 
\setlist[enumerate]{topsep=4pt,itemsep=3pt,parsep=0pt} 

\Crefname{figure}{Figure}{Figures}

\renewcommand{\int}{\mathrm{int}}

\newcommand{\N}[0]{\mathrm{\mathbb{N}}}

\newcommand{\tp}{\mathrm{tp}}

\newcommand{\dist}{\textnormal{dist}}

\renewcommand{\phi}{\varphi}

\newcommand{\XOR}{\textnormal{~~XOR~~}}

\newcommand{\Start}{\mathrm{start}}
\newcommand{\End}{\mathrm{end}}

\newcommand{\LL}{\mathcal{L}}

\newcommand{\DD}{\mathcal{D}}

\newcommand{\FO}{\mathrm{FO}}
\newcommand{\MSO}{\mathrm{MSO}}
\newcommand{\CMSO}{\mathrm{CMSO}}

\newcommand{\twins}{\mathrm{twins}}

\newcommand{\struc}[1]{\mathfrak{#1}}

\newcommand{\CC}{\mathcal{C}}

\newcommand{\FF}{\mathcal{F}}
\newcommand{\PP}{\mathcal{P}}
\newcommand{\QQ}{\mathcal{Q}}

\renewcommand{\SS}{\mathcal{S}}

\renewcommand{\le}{\leqslant}
\renewcommand{\leq}{\le}
\renewcommand{\ge}{\geqslant}
\renewcommand{\geq}{\ge}
\newcommand{\KK}{\mathcal{K}}

\newenvironment{claimproof}[1][\proofname]{%
  \begin{proof}[#1]%
}{%
  \end{proof}%
}

\begin{document}
\newcommand{\funding}{
The author was supported by the German Research Foundation (DFG) with grant agreement No. 444419611.
}

\title{Forbidden Induced Subgraphs for Bounded Shrub-Depth\\ and the Expressive Power of MSO\thanks{\funding}}
\date{}
\author{
  Nikolas M\"ahlmann \\
  \small{University of Bremen} \\
  \small{\texttt{maehlmann@uni-bremen.de}}
}
\maketitle

\pagenumbering{gobble}
\begin{abstract}
  The graph parameter shrub-depth is a dense analog of tree-depth.
  We characterize classes of bounded shrub-depth by forbidden induced subgraphs.
  The obstructions are well-controlled flips of large half-graphs and of disjoint unions of many long paths.
  Applying this characterization, we show that on every hereditary class of unbounded shrub-depth, MSO is more expressive than FO.
  This confirms a conjecture of 
  [Gajarsk{\'{y}} and Hlin{\v{e}}n{\'{y}}; LMCS 2015] who proved that on classes of bounded shrub-depth FO and MSO have the same expressive power.
  Combined, the two results fully characterize the hereditary classes on which FO and MSO coincide, answering an open question by [Elberfeld, Grohe, and Tantau; LICS 2012].

  Our work is inspired by the notion of stability from model theory.
  A graph class~$\CC$ is MSO-stable, if   no MSO-formula can define arbitrarily long linear orders in graphs from~$\CC$.
  We show that a hereditary graph class is MSO-stable if and only if it has bounded shrub-depth.
  As a key ingredient, we prove that every hereditary class of unbounded shrub-depth FO-interprets the class of all paths.
  This improves upon a result of [Ossona de Mendez, Pilipczuk, and Siebertz; Eur.\ J.\ Comb.\ 2025] who showed the same statement for FO-transductions instead of FO-interpretations.
\end{abstract}

\paragraph*{Acknowledgements.}
The author thanks Patrice Ossona de Mendez, Sebastian Siebertz, and Szymon Toru\'nczyk for fruitful discussions on the topic of this paper.

\newpage
\tableofcontents
\newpage
\pagenumbering{arabic}
\setcounter{page}{1}

\section{Introduction}
The main result of this paper is the following \cref{thm:main} which yields various characterizations for hereditary graph classes of bounded shrub-depth, in terms of forbidden induced subgraphs, monadic second-order logic~(MSO), counting monadic second-order logic~(CMSO), and first-order logic~(FO).
All notions appearing in \cref{thm:main} will be motivated, defined, and explained in the remainder of this introduction.

\begin{theorem}\label{thm:main}
  For every hereditary graph class~$\CC$, the following are equivalent.
  \begin{enumerate}
    \item $\CC$ has bounded shrub-depth.
    \item There is a~$t\in \N$ such that~$\CC$ excludes all flipped half-graphs of order~$t$ and
    all flipped~$tP_t$.
    \item There is a~$t\in \N$ such that~$\CC$ excludes all flipped half-graphs of order~$t$ and
    all flipped~$3P_t$.
    \item $\CC$ is MSO-stable.
    \item $\CC$ is monadically MSO-stable.
    \item $\CC$ is CMSO-stable.
    \item $\CC$ is monadically CMSO-stable.
    \item $\CC$ does not~$1$-dimensionally FO-interpret the class of all paths.
    \item FO and MSO have the same expressive power on~$\CC$.
  \end{enumerate}
\end{theorem}

\subsection*{Stability}
\begin{figure}[htbp]
  \centering
  \includegraphics[scale = 1]{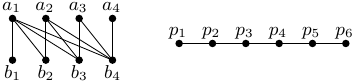}
  \caption{On the left: the half-graph of order~$4$ (denoted as~$H_4$) with vertices~$\{a_1,\ldots,a_4,b_1,\ldots,b_4\}$ and edges between~$a_i$ and~$b_j$ if~$i\leq j$.
  On the right: the~$6$-vertex path (denoted as~$P_6$).}
  \label{fig:h4-and-p5}
\end{figure}

We start motivating \cref{thm:main} by introducing the model-theoretic notion of \emph{stability}, a context in which the graph parameter shrub-depth will naturally arise.
Originating in the 70s and pioneered by Shelah, stability theory is a prolific branch of model theory which seeks to classify the complexity of theories (or in our case: graph classes).
There, stability is the most important dividing line, which separates the well-behaved \emph{stable} classes from the complex \emph{unstable} ones.
Intuitively, a graph class~$\CC$ is stable, if one cannot define arbitrarily large linear orders in~$\CC$ using logical formulas.
More precisely,
for a logic~$\LL\in\{\FO,\MSO, \CMSO\}$, an~$\LL$-formula~$\phi(\bar x,\bar y)$, and a graph class~$\CC$,
we say~$\phi$ has the \emph{order-property} on~$\CC$, if for every~$\ell \in \N$ there is a graph~$G\in\CC$ and a sequence~$\bar{a}_i, \ldots, \bar a_\ell$ 
of tuples of vertices of~$G$, such that
for all~$i,j\in[\ell]$
  \[G \models \phi(\bar{a}_i,\bar{a}_j) \quad \Leftrightarrow \quad  i \leq j.\]
For example the FO-formula~$\phi(x,y)$ expressing ``the neighborhood of~$x$ is a superset of the neighborhood of~$y$''
has the order-property on the class of all half-graphs, as it orders the sequence~$a_1,\ldots,a_t$ in the half-graph~$H_t$ of order~$t$ for every~$t\in\N$, as depicted in \cref{fig:h4-and-p5}.

Similarly, the MSO-formula
$\psi(x_1x_2,y_1y_2)$ expressing ``$x_1$ and~$x_2$ are not connected after deleting~$y_1$''
has the order-property on the class of all paths, as it orders the sequence of~$2$-tuples~$p_1p_t,p_2p_t,\ldots,p_tp_t$ in the~$t$-vertex path~$P_t$ for every~$t\in\N$.
A graph class~$\CC$ is \emph{$\LL$-stable} if no~$\LL$-formula has the order-property on~$\CC$, and \emph{$\LL$-unstable} otherwise.
The class of all half-graphs (all paths) is the arguably simplest example of an FO-unstable (MSO-unstable) graph class.

Apart from its extensive study on infinite structures in model theory, FO-stability has recently gained a lot of attention in finite model theory, in particular in structural and algorithmic graph theory. 
Podewski and Ziegler~\cite{stable_graphs} and Adler and Adler~\cite{adler2014interpreting} observed that on \emph{monotone}\footnote{A graph class is \emph{monotone}, if it is closed under taking subgraphs.} graph classes, FO-stability coincides with the combinatorial property of being \emph{nowhere dense}.
Nowhere dense classes are very general classes of sparse graphs~\cite{nevsetvril2011nowhere}, 
enjoying strong combinatorial and algorithmic properties, such as fixed-parameter tractable (fpt) FO model checking~\cite{grohe2017deciding}.
The equivalence of FO-stability and nowhere denseness in monotone classes elegantly bridges the fields of model theory and structural graph theory.
It has prompted the question whether also \emph{hereditary}\footnote{A graph class is \emph{hereditary}, if it is closed under taking induced subgraphs.} FO-stable graph classes are combinatorially well-behaved.
In the hereditary setting, FO-stability significantly generalizes nowhere denseness: for example the class of all cliques and the class of all map graphs are both FO-stable but not nowhere dense.
A current research program has uncovered multiple natural combinatorial characterizations of hereditary FO-stable classes~\cite{dreier2024stablemc, dreier2022indiscernibles, flipper-game, buffiere2024shallow} and shown that they also admit fpt FO model checking~\cite{flipper-game,dreier2024stablemc,dreier2023ssmc}.

In contrast to the rich literature on FO-stability, we are not aware of any previous works studying its natural restriction MSO-stability.
We attribute this to the fact that the compactness theorem, which is the main tool for working on \emph{infinite} structures where stability originates, fails for~MSO.
Now, the recent successes in the study of FO-stable classes of \emph{finite} graphs raise the question whether also MSO-stability can be understood through the lens of structural graph theory.
In this work we show that this is indeed the case, by proving the following.

\begin{theorem}\label{thm:mso-stable}
  A hereditary graph class is MSO-stable if and only if it has bounded shrub-depth.
\end{theorem}

The folklore fact that every monotone class of bounded shrub-depth also has bounded tree-depth yields the following corollary.

\begin{corollary}
  A monotone graph class is MSO-stable if and only if it has bounded tree-depth.
\end{corollary}

\subsection*{Shrub-Depth}
\emph{Shrub-depth} is a parameter for graph classes introduced in \cite{shrubdepth}.
It generalizes tree-depth to dense classes and can be seen as a bounded depth analog of clique- and rank-width, similar to how tree-depth is a bounded depth analog of tree-width.
Classes of bounded shrub-depth are extremely well-behaved from algorithmic, combinatorial, and logical points of view.
For instance, they admit quadratic time isomorphism testing \cite{ohlmann2023} and
fpt MSO model checking with an elementary dependence on the size of the input formula
\cite{gajarsky2015kernelizing,shrubdepth}; they can be characterized by vertex minors and FO-transductions \cite{shrubtrans}; and FO and MSO have the same expressive power on these classes \cite{gajarsky2015kernelizing}.

The above examples show that the structure side of bounded shrub-depth is well charted.
Indeed, we prove the direction ``bounded shrub-depth implies MSO-stability'' of \cref{thm:mso-stable} using mainly existing tools.
In contrast, proving the ``hereditary and unbounded shrub-depth implies MSO-instability'' direction requires insights about the non-structure side of shrub-depth (i.e., the properties shared by all classes that have unbounded shrub-depth).
Up until now, the picture here was much more vague:
In \cite{shrubdepth-journal} it is proved that for every~$d\in\N$ there exists a finite set of graphs~$\FF_d$ such that a graph has shrub-depth at most~$d$ if and only if it excludes all of~$\FF_d$ as induced subgraphs.
This result is non-constructive and does not reveal the concrete sets~$\FF_d$.
In \cite{KWON202176} and \cite{shrubtrans} it is shown that the class of all paths can be produced from any class~$\CC$ of unbounded shrub-depth by taking vertex-minors and also by an FO-transduction, respectively.
The high expressive power and irreversibility of vertex-minors and transductions limits our ability to draw conclusions about the structure of the original class~$\CC$ from these two results. 
In this paper, we improve the non-structure situation by providing a characterization of classes of bounded shrub-depth by explicitly listing forbidden induced subgraphs.

\subsection*{Forbidden Induced Subgraphs}
\begin{figure}[htbp]
  \centering
  \includegraphics{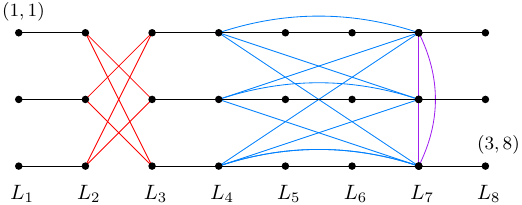}
  \caption{A flipped~$3P_8$. More precisely the depicted  graph is an~$\LL$-flip of~$3P_8$ with~$\LL = \{L_1,\ldots,L_8\}$, where the following parts were flipped:~$L_2$ with~$L_3$ (red),~$L_4$ with~$L_7$ (blue),~$L_7$ with~$L_7$ (purple).}
  \label{fig:flipped-3P9}
\end{figure}

\begin{figure}[htbp]
  \centering
  \includegraphics{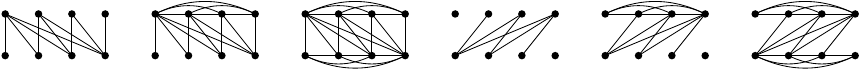}
  \caption{ 
  All flipped~$H_4$s (up to isomorphism).
  Figure replicated with permission from \cite[Fig.\ 2.5]{maehlmann-thesis}.
  }
  \label{fig:flipped-h4}
\end{figure}

We will briefly introduce the notions needed to state the obstructions.
For two graphs~$G$ and~$H$ on the same vertex set and a partition~$\PP$ of~$V(G)$, we say~$H$ is a \emph{$\PP$-flip} of~$G$ if it can be obtained by complementing the edge relation between pairs of parts of~$\PP$ (see \cref{fig:flipped-3P9}).
We refer to the preliminaries for formal definitions.
Our obstructions for bounded shrub-depth will be well-controlled flips of disjoint unions of long paths and of large half-graphs, as made precise in the following two definitions and their accompanying \cref{fig:flipped-3P9,fig:flipped-h4}.

\begin{restatable}{definition}{defSwimlanes}
  \label{def:tpt}
 ~$P_t$ is the~$t$-vertex path and~$mP_t$ is the disjoint union of~$m$ many~$P_t$, with vertices~$[m]\times [t]$, where~$(i,j)$ is the~$j$th vertex on the~$i$th path.
  A \emph{flipped}~$mP_t$ is an~$\LL$-flip of~$mP_t$ for the partition~$\LL = \{L_1,\ldots, L_t\}$ of the paths into \emph{layers}, where~$L_j = \{(1,j),\ldots,(m,j)\}$ contains the~$j$th vertices of all the paths for~$j\in[t]$.
\end{restatable}

\begin{restatable}{definition}{defFlippedHg}
  \label{def:hg}
  The \emph{half-graph of order~$t$} (denoted as~$H_t$) is the graph on vertices~$a_{1}, \ldots, a_{t}$ and~$b_{1}, \ldots, b_{t}$ where~$a_i$ and~$b_j$ are adjacent if and only if~$i \leq j$.
  A \emph{flipped}~$H_t$ is an~$\{A,B\}$-flip of ~$H_t$, where the flip-partition has parts~$A =\{a_{1}, \ldots, a_{t}\}$ and~$B = \{b_{1}, \ldots, b_{t}\}$.
\end{restatable}

We are now ready to state our characterization.

\begin{theorem}\label{thm:subgraphs}
  A graph class~$\CC$ has bounded shrub-depth if and only if there is~$t\in \N$ such that~$\CC$ excludes all flipped~$H_t$ and
  all flipped~$tP_t$ as induced subgraphs.
\end{theorem}

Moreover, our proofs show that in \cref{thm:subgraphs}, one can replace~$tP_t$ with~$3P_t$. 
While the stated~$tP_t$ variant of the theorem is more useful for hardness proofs, the~$3P_t$ variant is a step towards finding the simplest obstructions that cause unbounded shrub-depth.
It remains open whether the theorem also holds for~$2P_t$, but we know it fails for~$1P_t$, as every graph on~$t$ vertices is a flipped~$1P_t$.

\subsection*{Interpretations and Monadic Stability}
In order to prove the stability characterization (\cref{thm:mso-stable}) from the induced subgraph characterization (\cref{thm:subgraphs}), we visit another model-theoretic concept called \emph{interpretations}.
For a logic~$\LL$ a~$1$-dimensional\footnote{We will later also define and work with higher dimensional interpretations, whose formulas have tuples as free variables.}~$\LL$-interpretation~$I$ is defined by two~$\LL$-formulas: a domain formula~$\delta(x)$ and an irreflexive, symmetric edge formula~$\phi(x,y)$. 
It maps each graph~$G$ to the graph~$H := I(G)$ with vertex set~$V(H) := \{ v \in V(G) : G\models \delta(v) \}$ and edges~$E(H) := \{(u,v) \in V(H)^2 : G\models \phi(u,v)\}$.
We say a graph class~$\CC$ \emph{interprets} a graph class~$\DD$, if there is an interpretation~$I$ such that~$\DD \subseteq \{ I(G) : G\in \CC \}$.
Building on our forbidden induced subgraph characterization, we show the following.

\begin{theorem}\label{thm:interpreting-paths}
  A hereditary graph class has unbounded shrub-depth if and only if it~$1$-dimensionally FO-interprets the class of all paths.
\end{theorem}

As we have already seen, the class of all paths is MSO-unstable. \cref{thm:interpreting-paths} can be used to lift instability to all hereditary classes of unbounded shrub-depth.
This is a strengthening of a recent result by Ossona de Mendez, Pilipczuk, and Siebertz \cite{shrubtrans}, which states that every class of unbounded shrub-depth \emph{FO-transduces}\footnote{A transduction first colors the input graph non-deterministically, and then applies a fixed interpretation that can access these colors. For a single input graph~$G$, it produces multiple (possibly isomorphic) output graphs: one for each coloring of~$G$.
Due to the access to an arbitrary coloring, transductions are more expressive than interpretations.} the class of all paths.
Their result implies that every class of unbounded shrub-depth is \emph{monadically MSO-unstable}:
a graph class~$\CC$ is monadically~$\LL$-stable if every coloring of~$\CC$ is~$\LL$-stable.
In general, monadic MSO-stability is more restrictive than MSO-stability.
For instance the class~$\KK$ of all~$1$-subdivided cliques is MSO-stable, but monadically MSO-unstable.
When considering only hereditary classes, this example fails: the hereditary closure of~$\KK$ contains every~$1$-subdivided graph and is FO-unstable (so in particular also monadically FO-unstable, MSO-unstable, and monadically MSO-unstable).
Braunfeld and Laskowski showed that FO-stability and monadic FO-stability coincide in all hereditary classes~\cite{braunfeld2022existential}.
We show that the same collapse happens for MSO and CMSO.

\begin{theorem}
  For hereditary graph classes, the notions MSO-stability, monadic MSO-stability, CMSO-stability, and monadic CMSO-stability are all equivalent.
\end{theorem}

\subsection*{The Expressive Power of MSO}
For a graph class~$\CC$, we say FO and MSO \emph{have the same expressive power on~$\CC$} if for every MSO-sentence~$\phi$, there exists an FO-sentence~$\psi$ such that for every graph~$G\in\CC$ we have
$G \models \phi \Leftrightarrow G \models \psi$.
Otherwise, we say MSO is \emph{more expressive} than FO on~$\CC$.

It was shown by Grohe, Elberfeld, and Tantau that for all monotone classes~$\CC$, FO and MSO have the same expressive power if and only if~$\CC$ has bounded tree-depth~\cite{elberfeld2016}.
As an open question, they asked for a characterization of the hereditary classes where FO and MSO coincide.
As a dense analog of bounded tree-depth, bounded shrub-depth is a natural candidate here.
Indeed, Gajarsk\'{y} and Hlin\v{e}n\'{y} showed that
FO and MSO have the same expressive power on every class of bounded shrub-depth \cite[Thm.\ 5.14]{gajarsky2015kernelizing}.
However, they could not prove the reverse direction, which they attributed to a lack of known obstructions for classes of bounded shrub-depth.
As an application of our forbidden induces subgraph characterization, we provide the missing part of the puzzle by showing that MSO is more expressive than FO on every hereditary class of unbounded shrub-depth.
Together with the result by Gajarsk\'{y} and Hlin\v{e}n\'{y}, this completely characterizes the hereditary classes on which FO and MSO coincide.

\begin{theorem}
  For every hereditary graph class~$\CC$, FO and MSO have the same expressive power on~$\CC$ if and only if~$\CC$ has bounded shrub-depth.
\end{theorem}

\subsection*{Overview of the Paper}

\begin{figure}[ht]
\centering
\scalebox{0.75}{
\begin{tikzpicture}[%
  node distance=27mm,>=Latex,
  every edge/.style={draw=black,thin},
  block/.style={
    draw,
    fill=white,
    rectangle, 
    minimum width={2.5cm},
    minimum height={1.4cm},
    align = center
  }
  ]

\node[block] (pat1) 
{no flipped \\~$H_t$ and~$3P_t$};

\node[block, right = 1cm of pat1] (tpt) 
{no flipped \\~$H_t$ and~$tP_t$};

\node[block, right = 1.7cm of tpt] (ff) 
{$\infty$-flip-flat};

\node[block, right = 1cm of ff] (sd) 
{bounded \\ shrub-depth};

\node[block, below = 1cm of tpt] (coincide) 
{FO and MSO\\coincide};

\node[block, right = 1.7cm of sd] (mmsostable) 
{monadically \\ CMSO-stable};

\node[block, right = 1cm of mmsostable] (msostable) 
{MSO-stable};

\node[block, above = 1cm of tpt] (interp) 
{{not FO-} \\[-1mm] {interpreting} \\[-1mm] {all paths}};

\path[->] 
(tpt) edge node[above] {Prop.\ \ref{prop:flip-flat}} (ff);

\path[->] 
(sd.east) ++ (0,+1mm) edge node[above] {Prop.\ \ref{prop:sd-implies-mcmso-stable}} ([yshift=+1mm]mmsostable.west);

\path[<-] 
(sd.east) ++ (0,-1mm) edge node[below] {\cite{shrubtrans}} ([yshift=-1mm]mmsostable.west) ;

\path[->] 
(mmsostable) edge node[above] {def.} (msostable);

\draw[->] 
(msostable) |- node[above left=0cm and 7cm] {Lem.\ \ref{lem:unstable-interp}} (interp);

\path[<-]  
(tpt.north) edge node[right] {Prop.\ \ref{prop:patternsToPaths}} (interp.south);

\draw[<-]  (pat1.south)
-- node {} ++(0mm,-3.1cm) 
-| (mmsostable.south) node[pos=0.25, above] {Prop.\ \ref{prop:3ptUnstable}} node[pos=0.75] {};

\draw[<->] 
(ff) edge node[above] {\cite{dreier2024flipbreakability}} (sd);

\path[->] 
(coincide) edge node[right] {Prop.\ \ref{prop:sep-fo-mso}} (tpt);

\draw[<-] 
(coincide) -| node[above left=0cm and 3.5cm] {\cite{gajarsky2015kernelizing}} (sd);

\draw[->] 
(pat1) edge node[above] {def.} (tpt);

\end{tikzpicture}      
}
\caption{A map of \cref{thm:main}: combinatorial and logical characterizations of hereditary classes of bounded shrub-depth.}
\label{fig:implications}
\end{figure}
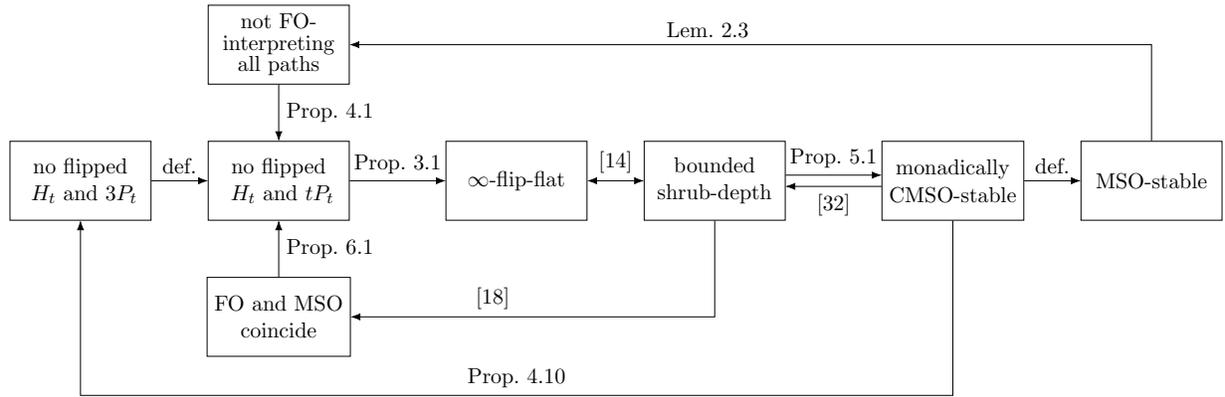

\cref{fig:implications} shows the implications which comprise \cref{thm:main}.
The only notion in the figure that has not been discussed so far is \emph{$\infty$-flip-flatness}.
This is another characterization of shrub-depth with strong ties to stability theory, recently proved by Dreier, Mählmann, and Toru\'{n}czyk~\cite{dreier2024flipbreakability},
which we define in the upcoming preliminaries (\cref{sec:prelims}).
The proofs of the individual implications are presented in \cref{sec:subgraphs,sec:interpreting,sec:mstable,sec:expressiveness}.
We conclude with an outlook in \cref{sec:outlook}, where we discuss potential strengthenings of our induced subgraph characterization and a second major dividing line from model theory, named \emph{dependence}.

\section{Preliminaries}\label{sec:prelims}
We use~$[n]$ to refer to the set~$\{ 1,\ldots,n \}$.
For an~$s$-tuple~$\bar a = a_1 \ldots a_s$, we use~$|\bar a|$ to refer to its length~$s$ and~$\bar a[i]$ to refer to its~$i$th component~$a_i$.

\subsection{Colored Structures and Graphs}
A \emph{(relational) signature}~$\Sigma$ is a set of relation symbols, each with an implicitly assigned arity.
A structure~$G$ over ~$\Sigma$ is called a \emph{$\Sigma$-structure}.
We denote by~$\Sigma^{(k)} := \Sigma \cup \{C_1,\ldots,C_k\}$ the \emph{expansion} of~$\Sigma$ by~$k$ many new unary predicates, which we can think of as colors.
A single element is allowed to have multiple colors, but this will not be important.
A~$\Sigma^{(k)}$-structure~$G^+$ is a \emph{$k$-coloring} of~$G$, if~$G$ is the~$\Sigma$-structure obtained by forgetting the color predicates in~$G^+$.
For a class of~$\Sigma$-structures~$\CC$, we denote by~$\CC^{(k)}$ the class of~$\Sigma^{(k)}$-structures consisting of all the~$k$-colorings of structures from~$\CC$.

A \emph{graph} is a~$\Gamma$-structure for the signature~$\Gamma := \{E\}$ consisting of a single binary relation (called \emph{edge relation}) that is interpreted symmetrically and irreflexive.
We call~$\Gamma^{(k)}$-structures \emph{$k$-colored graphs}.
For any structure~$G$, we use~$V(G)$ to refer to its universe: if~$G$ is a graph, then~$V(G)$ is its vertex set.

\subsection{Logic and Types}
We assume familiarity with first-order logic (FO) and monadic second-order logic (MSO).
As we model graphs as structures with a binary edge relation, MSO on graphs allows for quantification of vertex sets, but not of edge sets. This is also known as~$\MSO_1$ in the literature.
\emph{Counting} monadic second-order logic (CMSO) extends MSO for all set variables~$X$ and~$0 \leq r < k$ by the cardinality constraint~$\text{card}_{r,k}(X)$ that holds true if and only if~$|X| \equiv r \pmod k$.

Fix a logic~$\LL \in \{\FO,\MSO, \CMSO\}$ and a signature~$\Sigma$.
We use~$\LL[\Sigma]$ to refer to the set of~$\LL$-formulas over~$\Sigma$ and~$\LL[\Sigma]_q$
for the set of~$\LL$-formulas over~$\Sigma$ with quantifier rank at most~$q$.
We often drop the signature in the notation if it is clear from the context.
For a~$\Sigma$-structure~$G$, a tuple~$\bar a$, and a set of~$\Sigma$-formulas~$\Phi$, the \emph{type}~$\tp(\bar a, G, \Phi)$ is the set of formulas~$\phi(\bar x) \in \Phi$, with~$|\bar x| = |\bar a|$ such that~$G \models \phi(\bar a)$.
In particular~$\tp(G,\Phi)$ refers to the \emph{sentences}  (i.e., formulas without free variables) from~$\Phi$ that hold in~$G$. 
For example~$\tp(G,\FO_q)$ are the FO-sentences of quantifier rank at most~$q$ that hold in~$G$.

\subsection{Stability}
Fix a logic~$\LL$ and a signature~$\Sigma$.
An~$\LL[\Sigma]$-formula~$\phi(\bar x,\bar y)$ has the \emph{$\ell$-order-property} on a~$\Sigma$-structure~$G$, if there exists a sequence~$(\bar{a}_i)_{i \in [\ell]}$ 
of tuples of elements of~$G$, such that
for all~$i,j\in[\ell]$
  \[G \models \phi(\bar{a}_i,\bar{a}_j) \quad \Leftrightarrow \quad  i \leq j.\]
The formula~$\phi$ has the \emph{order-property} on a class of~$\Sigma$-structures~$\CC$, if  
for every~$\ell \in \N$ there is~$G \in \CC$ such that~$\phi$ has the~$\ell$-order-property on~$G$.
We call a class of~$\Sigma$-structures \emph{$\LL$-stable}, if no~$\LL[\Sigma]$-formula has the order-property on~$\CC$.
Moreover,~$\CC$ is \emph{monadically~$\LL$-stable} if for every~$k\in\N$, the class~$\CC^{(k)}$ of~$k$-colorings from~$\CC$ is~$\LL$-stable.

\subsection{Interpretations}

    For a logic~$\LL$,~$d\in\N$, and signatures~$\Sigma_1$ and~$\Sigma_2$,
    a \emph{$d$-dimensional~$\LL$-interpretation~$I$ from~$\Sigma_1$ to~$\Sigma_2$} consists of 
    \begin{itemize}
        \item an~$\LL[\Sigma_1]$-formula~$\delta_I(\bar x)$ called the \emph{domain formula},
        \item an~$\LL[\Sigma_1]$-formula~$\phi_{I,R}(\bar x_1,\ldots,\bar x_k)$ for every~$k$-ary relation~$R$ in~$\Sigma_2$,
    \end{itemize}
    where~$\bar x$ and all~$\bar x_i$ are~$d$-tuples.
    Given a~$\Sigma_1$-structure~$G$, we define the~$H := I(G)$ to be the~$\Sigma_2$-structure with the universe~$V(H) := \{\bar a \in V(G)^d : G \models \delta_I(\bar a)\}$
    and relations~$R(H) := \{ \bar a_1 \ldots \bar a_k \in V(H)^k : G \models \phi_{I,R}(\bar a_1,\ldots,\bar a_k) \}$ for every~$k$-ary relation~$R$ in~$\Sigma_2$.

We will use interpretations as a reduction mechanism.
The following lemma is crucial for this purpose. It is easy to prove by an inductive formula rewriting procedure. 
See, e.g., \cite[Thm.\ 4.3.1]{hodges-shorter}.

\begin{lemma}\label{lem:interp-fo}
    For every~$d$-dimensional~$\FO$-interpretation~$I$ from~$\Sigma_1$ to~$\Sigma_2$, and~$\FO[\Sigma_2]$-formula~$\phi(\bar x)$ there exists an~$\FO[\Sigma_1]$-formula~$\phi_I(\bar x)$ such that for every~$\Sigma_1$-structure~$G$ and tuple~$\bar a \in V(I(G))^{|\bar x|}$ we have~$I(G) \models \phi(\bar a)$ if and only if~$G \models \phi_I(\bar a)$.
\end{lemma}

The above lemma also holds for MSO- and CMSO-interpretations, when restricted to a single dimension.
See, e.g., \cite[Thm.\ 7.10]{courcelle2012graph}.
Higher dimensions are problematic because MSO and CMSO cannot quantify over sets of tuples. 

\begin{lemma}\label{lem:interp-mso}
  Fix~$\LL\in\{\MSO,\CMSO\}$.
  For every~$1$-dimensional~$\LL$-interpretation~$I$ from~$\Sigma_1$ to~$\Sigma_2$, and~$\LL[\Sigma_2]$-formula~$\phi(\bar x)$ there exists an~$\LL[\Sigma_1]$-formula~$\phi_I(\bar x)$ such that for every~$\Sigma_1$-structure~$G$ and tuple~$\bar a \in V(I(G))^{|\bar x|}$ we have~$I(G) \models \phi(\bar a)$ if and only if~$G \models \phi_I(\bar a)$.
\end{lemma}

For an interpretation~$I$ and a class~$\CC$, we write~$I(\CC)$ for the class~$\{ I(G) : G \in \CC \}$.
We say~$\CC$ \emph{$d$-dimensionally~$\LL$-interprets} a class~$\DD$, if there exists a~$d$-dimensional~$\LL$-interpretation~$I$ such that~$\DD \subseteq I(\CC)$.
We have already seen in the introduction, that the class of all paths is MSO-unstable. 
\cref{lem:interp-mso} lifts instability to all classes that~$1$-dimensionally MSO-interpret it.

\begin{lemma}\label{lem:unstable-interp}
  Fix~$\LL\in\{\MSO,\CMSO\}$.
  Every class that~$1$-dimensionally~$\LL$-interprets the class of all paths is~$\LL$-unstable.
\end{lemma}

\subsection{Transductions}
Intuitively, a transduction is the composition of a coloring step and an interpretation. We will make this more precise now.
Fix a logic~$\LL$ and signatures~$\Sigma_1, \Sigma_2$.
An~$\LL$-\emph{transduction~$T$ from~$\Sigma_1$ to~$\Sigma_2$} is defined by a~$1$-dimensional~$\LL$-interpretation~$I_T$ from~$\Sigma_1^{(k)}$ to~$\Sigma_2$ for some~$k \in \N$.
It maps a~$\Sigma_1$-structure~$G$ to the set of~$\Sigma_2$-structures~$T(G) := \{ I_T(G^+) : G^+ \text{ is a~$k$-coloring of~$G$} \}$.
For classes~$\CC$ and~$\DD$ we define~$T(\CC) := \bigcup_{G\in\CC} T(G)$ and
say~$\CC$ \emph{$\LL$-transduces~$\DD$} if there exists an~$\LL$-transduction~$T$ such that~$\DD \subseteq T(\CC)$.
Transductions play the role of interpretations in the monadic setting, and we have the following analog of \cref{lem:unstable-interp}.

\begin{lemma}\label{lem:unstable-trans}
    Fix~$\LL\in\{\MSO,\CMSO\}$.
    Every graph class that~$\LL$-transduces the class of all paths is monadically~$\LL$-unstable.
\end{lemma}

\subsection{Flips}
Fix a graph~$G$ and a partition~$\PP$ of its vertices.
For every vertex~$v \in V(G)$ we denote by~$\PP(v)$ the unique part~$Q \in \PP$ satisfying~$v \in Q$.
Let~$F \subseteq \PP^2$ be a symmetric relation.
We define~$H := G \oplus (\PP,F)$ to be the graph with vertex set~$V(G)$,
and edges defined by the following condition, for distinct~$u,v\in V(G)$:
\[
    uv \in E(H) \Leftrightarrow \begin{cases}
        uv \notin E(G) & \text{if } (\PP(u), \PP(v)) \in F,\\
        uv \in E(G) & \text{otherwise.}
    \end{cases}
\]
We call~$H$ a \emph{$\PP$-flip} of~$G$.
If~$\PP$ has at most~$k$ parts, we say that~$H$ a \emph{$k$-flip} of~$G$.
Flips are reversible:~$\big(G \oplus (\PP,F)\big) \oplus (\PP,F) = G$; and hereditary:
for every~$k$-flip~$H$ of~$G$ and~$A\subseteq V(G)$,
$H[A]$ is also a~$k$-flip of~$G[A]$.


It will be convenient to work with flip-partitions that are minimal in the following sense.

\begin{definition}
    Let~$H$ and~$G$ be two graphs on the same vertex set.
    We call a partition~$\PP$ of~$V(G)$ and a relation~$F \subseteq \PP^2$  \emph{irreducible~$(H,G)$-flip-witnesses} if
    \begin{itemize}
        \item $H = G \oplus (\PP,F)$,
        \item $H$ is not a~$(|\PP|-1)$-flip of~$G$,
        \item $F$ does not include~$(Q,Q)$ for any part~$Q \in \PP$ with~$|Q| = 1$.
    \end{itemize}
\end{definition}

Clearly for every two graphs~$H$ and~$G$ on the same vertex set, there exist irreducible~$(H,G)$-flip-witnesses.
In particular, as we work with loopless graphs, every flip relation~$F$ can be modified to satisfy the last condition stating that no singleton part is flipped with itself.
For the rest of this subsection, we argue that the irreducible flip-witnesses are uniquely determined.

\begin{lemma}\label{lem:discerning-part}
    Let~$H$ and~$G$ be two graphs on the same vertex set and~$\PP$ and~$F$ be  irreducible~$(H,G)$-witnesses.
    For every two distinct parts~$Q_1,Q_2 \in \PP$ there exists a part~$Q_\Delta \in \PP$ such that
    \[
        (Q_1,Q_\Delta) \in F \Leftrightarrow  (Q_2,Q_\Delta) \notin F.
    \]
    We say~$Q_\Delta$ \emph{discerns}~$Q_1$ and~$Q_2$.
\end{lemma}
\begin{proof}
    If no part discerns~$Q_1$ and~$Q_2$, we can merge the two parts and obtain a smaller partition; a contradiction to irreducibility.
\end{proof}

\begin{lemma}\label{lem:coarsening}
    Let~$H$ and~$G$ be two graphs on the same vertex set, let~$\PP$ and~$F$ be irreducible~$(H,G)$-flip-witnesses, and let~$\PP'$ and~$F'$ any partition and relation such that~$H = G \oplus (\PP',F')$.
    Then~$\PP$ is a coarsening of~$\PP'$.
\end{lemma}
\begin{proof}
    Assume towards a contradiction the existence of two vertices~$u$ and~$v$ with~$u \in Q_1$ and~$v \in Q_2$ but~$u,v \in Q'$ for distinct parts~$Q_1,Q_2 \in \PP$ and a part~$Q' \in \PP'$.
    By \cref{lem:discerning-part}, there exists a part~$Q_\Delta \in \PP$ discerning~$Q_1$ and~$Q_2$.

    Assume first there exists a discerning part~$Q_\Delta$ containing a vertex~$w \notin \{u,v\}$. 
    If both~$u$ and~$v$ have the same adjacency towards~$w$ in~$G$, 
    they still have the same adjacency towards~$w$ in~$G \oplus (\PP',F')$, as they are in the same part~$Q' \in \PP'$, but differing adjacencies towards~$w$ in~$G \oplus (\PP,F)$.
    If~$u$ and~$v$ have differing adjacencies towards~$w$ in~$G$, 
    they still have differing adjacencies towards~$w$ in~$G \oplus (\PP',F')$, but the same adjacencies towards~$w$ in~$G \oplus (\PP,F)$.
    In both cases we have a contradiction to~$G \oplus (\PP,F) = G \oplus (\PP',F')$.    

    Assume now every discerning part~$Q_\Delta$ is either~$Q_\Delta = \{u\}$ or~$Q_\Delta = \{v\}$.
    Up to symmetry, we have that~$Q_\Delta = \{u\} = Q_1$ is discerning.
    Now if also~$Q_2$ is a singleton part~$Q_2 = \{v\}$, then we can merge~$Q_1$ and~$Q_2$ to obtain a partition witnessing that~$H$ is a~$(|\PP|-1)$-flip of~$G$; a contradiction.
    Therefore,~$Q_2$ has at least two elements and is not discerning. 
    As singleton parts are not flipped with itself and as~$Q_\Delta$ is discerning, we know that~$(Q_1,Q_1 = Q_\Delta) \notin F$ and~$(Q_2, Q_1 = Q_\Delta) \in F$.
    As~$Q_2$ is not discerning, we must have also~$(Q_2, Q_2) \in F$.
    But if both~$(Q_1 = \{u\},Q_2)$ and~$(Q_2,Q_2)$ are in~$F$ and no part other than~$Q_1$ discerns~$Q_1$ and~$Q_2$, then we can again merge~$Q_1$ and~$Q_2$ to obtain a partition witnessing that~$H$ is a~$(|\PP|-1)$-flip of~$G$; a contradiction.
\end{proof}

\begin{lemma}\label{lem:irreducible-unique}
    For every two graphs~$H$ and~$G$ on the same vertex set, the irreducible~$(H,G)$-flip-witnesses are uniquely determined.
\end{lemma}
\begin{proof}
    Assume towards a contradiction the existence of two different irreducible flip-witnesses~$\PP$ and~$F$, and~$\PP'$ and~$F'$.
    If~$\PP = \PP'$ then also~$F = F'$,
    so assume~$\PP \neq \PP'$.
    Up to symmetry, there exist two vertices~$u$ and~$v$ with~$u \in Q_1$ and~$v \in Q_2$ but~$u,v \in Q'$ for distinct parts~$Q_1,Q_2 \in \PP$ and a part~$Q' \in \PP'$.
    By \cref{lem:coarsening}, we know that~$\PP'$ is a coarsening of~$\PP$.
    This means~$\PP'$ has strictly fewer parts than~$\PP$; a contradiction to~$\PP$ irreducible.
\end{proof}

For every two graph~$H$ and~$G$ on the same vertex set, we call~$\PP$ the \emph{irreducible~$(H,G)$-flip-partition} and~$F$ the \emph{irreducible~$(H,G)$-flip-relation}, if~$\PP$ and~$F$ are the unique irreducible~$(H,G)$-flip-witnesses, as justified by \cref{lem:irreducible-unique}.

\subsection{Shrub-Depth and SC-Depth}
\emph{Shrub-depth}~\cite{shrubdepth,shrubdepth-journal} is a parameter for graph classes.
Unlike for example tree-width, one cannot meaningfully measure the shrub-depth of a single graph.
In this paper we will not define shrub-depth, but instead work with the functionally equivalent \emph{SC-depth}, which was introduced in~\cite{shrubdepth} and is definable for single graphs.
We first need to define the eponymous notion of a \emph{set-complementation}.
A graph~$H$ is a set complementation of a graph~$G$, if~$H$ can be obtained from~$G$ by complementing the edges on a subset of vertices. In the language of flips:
$H = G \oplus (\{A, V(G)\setminus A\}, \{(A,A)\})$ for some set~$A\subseteq V(G)$.

The single vertex graph~$K_1$ has SC-depth~$0$.
A graph has SC-depth at most~$d$ if it is a set-complementation of a disjoint union of (arbitrarily many) graphs of SC-depth at most~$d-1$.
The SC-depth of a graph~$G$ is the smallest value~$d$ such that~$G$ has SC-depth at most~$d$.
A graph class~$\CC$ has \emph{bounded SC-depth} if there is~$d\in\N$ such that every graph in~$\CC$ has SC-depth at most~$d$.

\begin{theorem}[\cite{shrubdepth}]\label{thm:sc-depth}
    A graph class has bounded SC-depth if and only if it has bounded shrub-depth.
\end{theorem}

\subsection{Distances and Flatness}
The \emph{length} of a path is the number of edges it contains, i.e.,~$P_t$ has length~$t-1$.
For a graph~$G$ and two vertices~$u,v \in V(G)$ we define~$\dist_G(u,v)$ to be the length of a shortest path between~$u$ and~$v$ in~$G$.
If no such path exists, then~$u$ and~$v$ are in different connected components of~$G$, and we define~$\dist_G(u,v) := \infty$.
For~$r\in\N$, a graph~$G$, and a vertex~$v \in V(G)$ let~$N_r^G[v] := \{u \in V(G) : \dist_G(u,v) \leq r\}$ be the \emph{(closed) radius-$r$ neighborhood of~$v$}.
We drop the superscript~$G$ if it is clear from the context.
For~$r\in\N$, we call a set~$A \subseteq V(G)$ \emph{distance-$r$ independent}  in~$G$ if~$\dist_G(u,v) > r$ for every two distinct vertices~$u,v \in A$.
Similarly,~$A$ is \emph{distance-$\infty$ independent} in~$G$ if no two vertices of~$A$ are in the same connected component of~$G$.
\begin{definition}
    For~$r\in \N \cup \{\infty\}$ and~$k\in\N$, a set~$A$ of vertices in a graph is \emph{$(r,k)$-flip-flat}, if there exists a~$k$-flip~$H$ of~$G$ in which~$A$ is distance-$r$ independent.
    A graph class~$\CC$ is 
    \emph{$r$-flip-flat},
        if there are \emph{margins}~$k_r\in\N$ and~$M_r:\N\to\N$, such that for every~$G\in\CC$ and~$m\in\N$, every set~$W \subseteq V(G)$ of size~$|W|\geq M_r(m)$ contains an~$(r,k_r)$-flip-flat subset of size~$m$.
\end{definition}

We refer to \cite{maehlmann-thesis} for an introduction to flatness properties.
Crucially, flip-flatness characterizes both monadic FO-stability~\cite{dreier2022indiscernibles} and bounded shrub-depth~\cite{dreier2024flipbreakability}.

\begin{theorem}[{\cite{dreier2022indiscernibles} and \cite{dreier2024flipbreakability}}]\label{thm:flip-flat}
    For every graph class~$\CC$:
    \begin{itemize}
        \item $\CC$ is FO-stable if and only if~$\CC$ is~$r$-flip-flat for every~$r\in\N$.
        \item $\CC$ has bounded shrub-depth if and only if~$\CC$ is~$\infty$-flip-flat.
    \end{itemize}
\end{theorem}



\section{Forbidden Induced Subgraphs}\label{sec:subgraphs}

Flipped half-graphs~$H_t$ and flipped~$tP_t$s were defined in the introduction (\cref{def:hg,def:tpt}).
We say a graph class~$\CC$ is \emph{pattern-free}, if there is a~$t\in \N$ such that~$\CC$ excludes as induced subgraphs every flipped~$H_t$ and
every flipped~$tP_t$.
The goal of this section is to show the following.

\begin{restatable}{proposition}{propFlipFlat}
  \label{prop:flip-flat}
  Every pattern-free class is~$\infty$-flip-flat.
\end{restatable}

Before we dive into the proof of \cref{prop:flip-flat}, let us quickly sketch an argument showing the reverse direction: every~$\infty$-flip-flat class is pattern-free.
We will be brief here, as this direction will later also be implied by our other proofs (see \cref{fig:implications}).
By \cref{thm:flip-flat,thm:sc-depth}, it suffices to show that every class that is not pattern-free has unbounded SC-depth.
It is well known that (flipped) half-graphs have unbounded SC-depth.
To show the same for flipped~$tP_t$s,
it suffices to prove the following two lemmas. Together they imply that any SC-depth decomposition of a flipped~$tP_t$ has a branch whose depth is unbounded in~$t$.

\begin{lemma}
  Every set-complementation of a flipped~$sP_t$ contains a flipped~$tP_t$ for~$s := t \cdot 2^t$.
\end{lemma}
\begin{proof}
  Let~$G$ be a set-complementation of a flipped~$sP_t$ where the set~$A \subseteq V(sP_t)$ was complemented.
  By the pigeonhole principle, there exist at least~$t$ among the~$s$ many~$P_t$ whose vertices have the same membership in~$A$. This means for every~$i \in [t]$  the~$i$th vertices of all the paths are either all in~$A$ or all not in~$A$.
  These~$t$ many paths form an induced flipped~$tP_t$.
\end{proof}

\begin{lemma}
  Every flipped~$tP_s$ has a connected component containing a flipped~$tP_t$ for~$s := t^2 + t - 1$.
\end{lemma}
\begin{proof}
  Let~$G$ be a flipped~$tP_s$.
  If at least a single flip between any two layers was performed, then~$G$ is connected, and we are done.
  Otherwise, no flip was performed and~$G = tP_s$ has~$P_s$ as a connected component.
  Cutting~$P_s$ into~$t$ many~$P_t$s yields an induced~$tP_t$.
\end{proof}

\subsection{Establishing Monadic FO-Stability}

As a first step, we will show that pattern-free classes are monadically FO-stable.
This will follow from a known characterization of monadic FO-stability by forbidden induced subgraphs.
We first introduce the required definitions.

\begin{figure}[htbp]
  \centering
  \includegraphics[scale = 0.7]{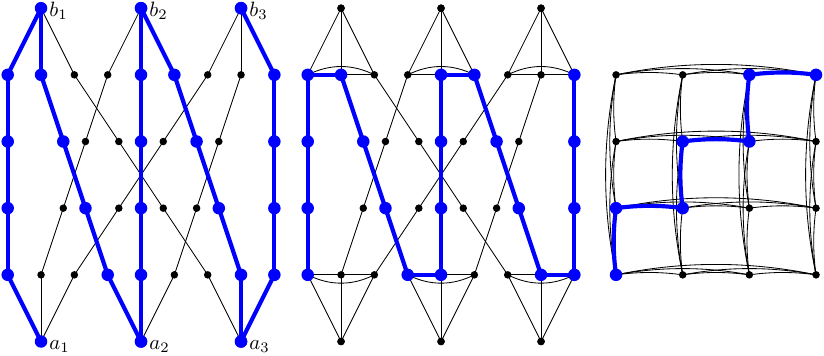}
  \caption{From left to right: the star 4-crossing of order~$3$,
  the clique 4-crossing of order~$3$, and the rook graph of order~$4$.
  Highlighted in blue we show how each of the three patterns of order~$k$ contains an induced path on at least~$k$ vertices.
  }
  \label{fig:fo-patterns}
\end{figure}

For~$r \ge 1$, the \emph{star~$r$-crossing} of order~$t$ is the~$r$-subdivision of~$K_{t,t}$ (the biclique of order~$t$).
More precisely, it consists of \emph{roots}~$a_1,\dots,a_t$ and~$b_1,\dots,b_t$
together with~$t^2$ many pairwise vertex-disjoint~$r$-vertex paths~$\{ \pi_{i,j} : i,j \in [t] \}$, whose endpoints we denote as~$\Start(\pi_{i,j})$ and~$\End(\pi_{i,j})$.
Each root~$a_i$ is adjacent to~$\{ \Start(\pi_{i,j}) : j \in [t] \}$,
and each root~$b_j$ is adjacent to~$\{ \End(\pi_{i,j}) : i \in [t] \}$.
See \Cref{fig:fo-patterns}.
The \emph{clique~$r$-crossing} of order~$t$ is the graph obtained from the star~$r$-crossing of order~$t$
by turning the neighborhood of each root into a clique.
In order to define \emph{flipped} versions of star/clique~$r$-crossings, we partition their vertices into \emph{layers}~$\LL = \{L_0, \ldots, L_{r+1}\}$:
The 0th layer consists of the vertices~$\{a_1,\dots,a_t\}$.
The~$l$th layer, for~$l \in [r]$, consists of the~$l$th vertices of the paths~$\{ \pi_{i,j} : i,j \in [t] \}$.
Finally, the~$(r+1)$th layer consists of the vertices~$\{b_1,\dots,b_n\}$. 
A \emph{flipped} star/clique~$r$-crossing
is a graph obtained from a 
star/clique~$r$-crossing
by performing a flip where the parts of the flip-partition are the layers of the~$r$-crossing.
The following characterization was originally proven in \cite{dreier2024stablemc}, but we refer to \cite{maehlmann-thesis} for this formulation.

\begin{theorem}[{\cite{dreier2024stablemc}}]\label{lem:fo-patterns}
  A graph class is monadically FO-stable if and only if
  for every~$r \geq 1$ there exists~$t \in \N$
        such that~$\CC$ excludes as induced subgraphs
        \begin{itemize}
            \item all flipped star~$r$-crossings of order~$t$, and
            \item all flipped clique~$r$-crossings of order~$t$, and
            \item all flipped~$H_t$.
        \end{itemize}
\end{theorem}

\begin{lemma}\label{lem:crossing-to-path}
  For every~$r\geq 1$ and~$t \in \N$, the star~$r$-crossing and the clique~$r$-crossing of order~$t$ both contain an induced~$P_t$.
\end{lemma}
\begin{proof}
  For star~$r$-crossings with~$r \geq 1$ and for clique~$r$-crossing with~$r\geq 2$, this is easy to see in the left/middle panel of \cref{fig:fo-patterns}.
  For the clique~$1$-crossing of order~$t$, note that its vertices~$\{\Start(\pi_{i,j}) = \End(\pi_{i,j}) : i,j \in [t]\}$ induce the \emph{rook graph} of order~$t$. 
  This graph is defined as the graph on vertices~$[t] \times [t]$, where the vertices~$(i,j)$ and~$(i',j')$ are adjacent if and only if~$i = i'$ or~$j = j'$.
  The right panel of \cref{fig:fo-patterns} shows how~$P_t$ embeds into the rook graph of order~$t$.
\end{proof}

\begin{lemma}\label{lem:fo-unstable-to-ideal-patterns}
  For every monadically FO-unstable graph class~$\CC$ there exists~$k\in\N$ such that~$\CC$ contains as induced subgraphs either
  \begin{itemize}
    \item a flipped~$H_t$ for every~$t\in\N$, or
    \item a~$k$-flip of~$P_t$ for every~$t \in \N$.
  \end{itemize}
\end{lemma}
\begin{proof}
  By \cref{lem:fo-patterns} and the pigeonhole principle, there exists some~$r \in \N$ such that~$\CC$ contains as induced subgraphs either
  \begin{enumerate}
    \item a flipped~$H_t$ for every~$t\in\N$, or
    \item a flipped star~$r$-crossing of order~$t$ for every~$t\in\N$, or
    \item a flipped clique~$r$-crossing of order~$t$ for every~$t\in\N$.
  \end{enumerate}
  In the first case we are done.
  In the other two cases we conclude by observing that every flipped~$r$-crossing is in particular a~$k$-flip of an~$r$-crossing for~$k := r +2$ and by applying \cref{lem:crossing-to-path}.
\end{proof}

\begin{lemma}\label{lem:pigeonhole-swimlane}
  Every~$k$-flip of~$(m \cdot k^t)P_t$ contains an induced flipped~$mP_t$, for every~$k,m,t \in \N$.
\end{lemma}
\begin{proof}
  Choose any~$k$-flip of~$(m \cdot k^t)P_t$ and let~$\PP$ be a witnessing partition of size at most~$k$.
  We can color the vertices of the~$P_t$ according to their membership in~$\PP$. 
  This yields at most~$k^t$ different ways to color a~$P_t$.
  By the pigeonhole principle, there must be a set~$\SS$ of at least~$m$ many~$P_t$ that are colored in the same way.
  This means for every~$i \in [t]$ all the~$i$th vertices of paths in~$\SS$ are in the same part of~$\PP$.
  Hence, the paths in~$\SS$ induce a flipped~$mP_t$.
\end{proof}

By cutting a long path into multiple shorter ones, we observe that~$P_{t'}$ contains an induced~$mP_t$ for~$t' := m \cdot (t + 1)$. 
Combining this observation with \cref{lem:pigeonhole-swimlane} yields the following corollary.

\begin{corollary}\label{lem:path-to-swimlane}
  Every~$k$-flip of~$P_{t'}$ contains an induced flipped~$tP_t$ for
  every~$k,t \in\N$ and~$t':= t \cdot k^t \cdot (t+1)$.
\end{corollary}

Combining this with \cref{lem:fo-unstable-to-ideal-patterns} yields the following.

\begin{proposition}\label{prop:pattern-free-fo-stable}
  Every pattern-free class is monadically FO-stable.
\end{proposition}

\subsection{Proving \texorpdfstring{$\infty$-Flip-Flatness}{infty-Flip-Flatness}}

\begin{lemma}\label{lem:flat-or-pattern}
  For every~$m,t\in\N$, graph~$G$, and distance-$2t$ independent set~$A$ of size~$2m$ in~$G$ either
  \begin{itemize}
    \item a size~$m$ subset of~$A$ is distance-$\infty$ independent in~$G$, or
    \item $G$ contains an induced~$mP_t$.
  \end{itemize}
  
\end{lemma}

\begin{figure}[htbp]
  \centering
  \includegraphics[width = \textwidth]{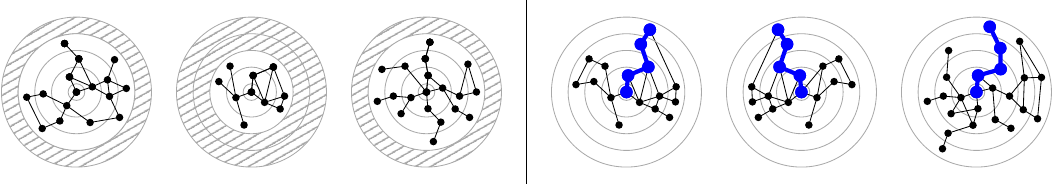}
  \caption{The two cases from \cref{lem:flat-or-pattern}.
  The circles are the layering of a BFS tree around the vertices of~$A$.
  There is a large subset~$A'\subseteq A$ whose outermost layers are either all empty (left panel), or all non-empty (right panel).
  Therefore, the vertices of~$A'$ are either in pairwise different connected components, or the vertices of~$A'$ are the endpoints of pairwise vertex-disjoint long induced paths.
  }
  \label{fig:flat-or-pattern}
\end{figure}

\begin{proof}
  By the assumption of the lemma, the radius-$(t-1)$ neighborhoods of the vertices in~$A$ are pairwise disjoint, and every edge is incident to at most one of these neighborhoods.
  For each~$a \in A$, let~$X(a) \subseteq N_{t-1}[a]$ be the set of vertices that are at distance exactly~$t-1$ from~$a$ in~$G$.
  By the pigeonhole principle, there is a set~$A' \subseteq A$ of size~$m$ such that either~$X(a)$ is empty for all~$a \in A'$ or~$X(a)$ is non-empty for all~$a \in A'$ (see \cref{fig:flat-or-pattern}).
  In the first case this means for each~$a \in A'$,~$N_{t-1}[a]$ contains the entire connected component of~$a$ in~$G$,
  and~$A'$ is distance-$\infty$ independent.
  In the second case for each~$a \in A'$,~$N_{t-1}[a]$ contains an induced~$P_t$ (which starts at~$a$ and ends in~$X(a)$), and~$G$ contains an induced~$mP_t$.
\end{proof}

Using \cref{lem:pigeonhole-swimlane}, the previous lemma lifts to flipped graphs.

\begin{corollary}\label{cor:infty-flip-flat-engine}
  For every~$k,m,t \in \N$, graph 
 ~$G$,~$k$-flip~$H$ of~$G$, and distance-$2t$ independent set~$A$ of size~$2k^t \cdot m$ in~$H$ either 
  \begin{itemize}
    \item a size~$k^t \cdot m$ subset of~$A$ is distance-$\infty$ independent in~$H$, or
    \item $G$ contains an induced flipped~$mP_t$.
  \end{itemize}
\end{corollary}

We are now ready to prove \cref{prop:flip-flat}, which we restate for convenience.

\propFlipFlat*

\begin{proof}
  Fix a pattern-free class~$\CC$.
  Then there is~$t\in \N$ such that~$\CC$ contains no induced flipped~$tP_t$.
  By \cref{prop:pattern-free-fo-stable},~$\CC$ is monadically FO-stable, and by \cref{thm:flip-flat}, also~$r$-flip-flat 
  for some margins~$k_r$ and~$M_r$ for every~$r\in\N$.
  We claim that~$\CC$ is~$\infty$-flip-flat with margins~$k_\infty := k_{2t}$ and~$M_\infty(m) := M_{2t}(\max(2m,2k_{2t}^t \cdot t))$.
  Choose any~$G\in \CC$,~$m\in \N$, and~$W \subseteq V(G)$ of size~$|W| \geq M_\infty(m)$.
  By flip-flatness, there exists a~$k_\infty$-flip~$H$ of~$G$ in which a size~$\max(2m,2k_{2t}^t \cdot t)$ subset~$A \subseteq W$ is distance-$2t$ independent.
  By \cref{cor:infty-flip-flat-engine}, either
  \begin{itemize}
    \item a size~$\max(2m,2k_{2t}^t \cdot t)/2 \geq m$ subset of~$A \subseteq W$ is distance-$\infty$ independent in~$H$, or
    \item $G$ contains an induced flipped~$m'P_t$ for some~$m' \geq \max(2m,2k_{2t}^t \cdot t) / (2k^t_{2t}) \geq t$.
  \end{itemize}
  The second case contradicts pattern-freeness of~$\CC$, so we have successfully established~$\infty$-flip-flatness for~$\CC$.
\end{proof}

\section{Interpreting Paths}\label{sec:interpreting}

In this section we will prove the following.
\begin{restatable}{proposition}{propPatternsToPaths}
\label{prop:patternsToPaths}
    There is a~$1$-dimensional FO-interpretation~$I$ such that for every hereditary graph class~$\CC$, that for every~$t\in\N$ contains either a flipped~$H_t$ or a flipped~$tP_t$,
    the class of all paths is contained in~$I(\CC)$.
\end{restatable}

Notably, a single interpretation works for all classes that contain these patterns.
We first show that the paths of length at most~$3$ appear already as induced subgraphs.

\begin{lemma}\label{lem:p3-from-swimlane}
    Every flipped~$2P_{3}$ and every flipped~$H_3$ contains~$P_3$ as an induced subgraph.
\end{lemma}

\begin{figure}[htbp]
    \centering
    \includegraphics[width = \textwidth]{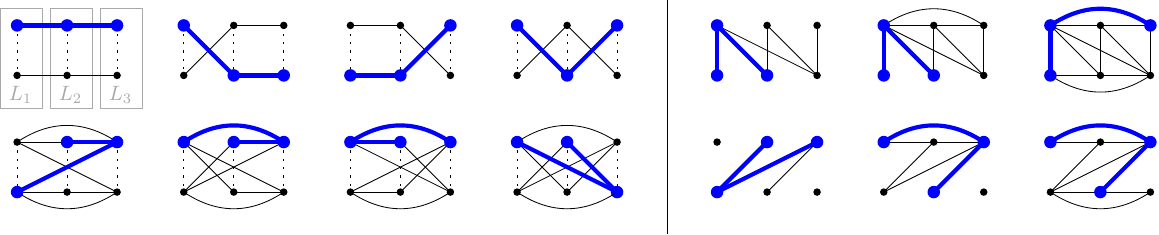}
    \caption{On the left: An enumeration of all flipped~$2P_3$.
    To reduce the number of cases, we do not account for flips that flip a layer~$L_i$ with itself. Edges that could possibly be created by such flips are marked as dashed.
    In each flipped~$2P_3$ we have highlighted an induced~$P_3$ that contains no dashed edges.
    This proves that every flipped~$2P_3$ contains an induced~$P_3$.
    On the right: An enumeration of all flipped~$H_3$, each with a highlighted induced~$P_3$.
    }
    \label{fig:p3}
\end{figure}

\begin{proof}
    Exhaustive case distinctions are depicted in \cref{fig:p3}.
\end{proof}

\subsection{Interpreting Paths in Flipped \texorpdfstring{$tP_t$s}{tPts}}

Two vertices~$u$ and~$v$ are \emph{twins} in a graph~$G$, if~$N^G_1[u]\setminus \{ u,v \} = N^G_1[v]\setminus \{ u,v \}$.
This relation is FO-definable:
\[
    \twins(x,y) := \forall z : (z \neq x \wedge z \neq y) \rightarrow (E(z,x) \leftrightarrow E(z,y)).
\]

\begin{observation}\label{obs:twin-flip}
    Let~$H$ be a~$\PP$-flip of~$G$. Two vertices~$u,v \in V(G)$ that are contained in the same part of~$\PP$ are twins in~$G$ if and only if they are twins in~$H$.
\end{observation}

\begin{lemma}\label{lem:paths-from-swimlanes}
    There exists an FO-interpretation~$I$ such that for every~$t\geq 4$, every flipped~$5P_t$ contains an induced subgraph~$H$ such that~$I(H)= P_t$.

    \smallskip\noindent
    Moreover,~$H$ contains at least~$8$ vertices that have a twin in~$H$.
\end{lemma}

The ``moreover'' part of the lemma is later used to construct a single interpretation that can distinguish whether the input graph is an induced subgraph of a flipped~$tP_t$ or of a flipped~$H_t$.

\begin{figure}[htbp]
    \centering
    \includegraphics{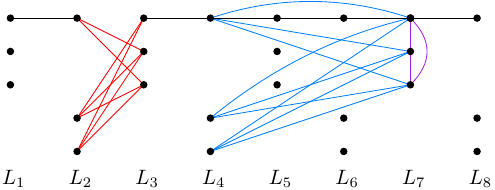}    
    \caption{The induced subgraph of a flipped~$5P_8$ in which the interpretation~$I$ can interpret~$P_8$.
    In this example the following layers were flipped:~$L_2$ with~$L_3$ (red),~$L_4$ with~$L_7$ (blue),~$L_7$ with~$L_7$ (purple).
    }
    \label{fig:pt-from-flipped-5pt}
\end{figure}

\begin{proof}[Proof of \cref{lem:paths-from-swimlanes}]
    By assumption~$\CC$ contains a graph~$H'$ that is an~$\LL$-flip of~$5P_t$, where~$\LL$ is the usual partition of the vertices of~$5P_t$ into layers.
    We will interpret~$P_t$ in the induced subgraph~$H := H'[A]$ where~$A := X \cup S$ with
    \[
        X = \{(1,i) : i \in [t]\} 
        \quad\text{and}\quad 
        S := S_1 \cup \dots \cup S_t
        \quad\text{and}\quad 
        S_i := 
        \begin{cases}
            \{(2,i),(3,i)\} & \text{if~$i$ is odd,}\\
            \{(4,i),(5,i)\} & \text{if~$i$ is even}. 
        \end{cases}
    \]
    See \cref{fig:pt-from-flipped-5pt} for an example.
    In the non-flipped setting, taking the subgraph of~$5P_t$ induced on the same vertex set~$A$ yields a graph~$G:=(5P_t)[A]$ that is the disjoint union of a single~$P_t$ (formed by the~$X$ vertices) and an independent set of size~$2t$ (formed by the~$S$ vertices).
    Observe that~$H$ is an~$\LL\vert_A$-flip of~$G$, where~$\LL\vert_A$ is the restriction of~$\LL$ to~$A$.
    Let~$\QQ$ and~$F \subseteq \QQ^2$ be the irreducible~$(H,G)$-flip-witnesses such that~$H = G \oplus (\QQ,F)$.
    By \cref{lem:coarsening},~$\QQ$ is a coarsening of~$\LL\vert_A$ and, by \cref{lem:discerning-part}, for every two distinct parts~$Q_1,Q_2 \in \QQ$ there exists a discerning part~$Q_\Delta \in \QQ$ such that 
    \[
        (Q_1,Q_\Delta) \in F \Leftrightarrow  (Q_2,Q_\Delta) \notin F.
    \]
    In order to interpret~$P_t$ from~$H$, it suffices to establish the following two claims.
    
    \begin{claim}\label{clm:domain}
        The set~$X$ is definable in~$H$. This means there is a formula~$\delta(x)$ such that for all~$v \in V(H) = A$ we have
        \[
          H \models \delta(v) \Leftrightarrow v \in X.  
        \]
    \end{claim}
    
    \begin{claim}\label{clm:edges}
        The edges of~$G[X]$ are definable in~$H$. This means there is a formula~$\phi(x,y)$ such that for all~$u,v \in X$ we have
        \[
            H \models \phi(u,v) \Leftrightarrow G \models E(u,v).  
        \]
    \end{claim}

    If we define the interpretation~$I$ using the formulas~$\delta_I := \delta$ and~$\phi_{I,E} := \phi$ from \cref{clm:domain} and \cref{clm:edges}, then~$I(H) = P_t$.
    It will be clear from the proofs of the two claims, that we can choose~$\delta$ and~$\phi$ independently of~$t$, such that~$I$ is the desired interpretation.
    \begin{claimproof}[Proof of \cref{clm:domain}]
        We choose~$\delta(x) := \forall y: x \neq y \rightarrow \neg \twins(x,y)$ to be the formula expressing that~$x$ has no twins in~$H$.
        We first verify that no vertex in~$S = A \setminus X$ satisfies~$\delta$.
        Indeed, each vertex~$u \in S_i \in S$ has a twin~$v$ in~$S_i$ in the graph~$G$: both vertices have no neighbors at all in~$G$.
        As both are in the same part of~$\LL|_A$ and its coarsening~$\QQ$, they are also twins in~$H$ by \cref{obs:twin-flip}.
        
        We next verify that every vertex in~$X$ satisfies~$\delta$.
        Fix a vertex~$u \in X$ and let~$Q_1$ be the part of~$\QQ$ containing it.
        Since we assumed~$t\geq 4$, we know that~$u$ has no twins in~$G$.
        Then, again by \cref{obs:twin-flip},~$u$ has no twins in~$H$ that are also contained in the same part~$Q_1$.
        Now let~$v \in Q_2 \in \QQ$ be a vertex contained in a different part.
        By \cref{lem:discerning-part}, there exists a part~$Q_\Delta$ such that~$(Q_1,Q_\Delta) \in F \Leftrightarrow  (Q_2,Q_\Delta) \notin F$.
        This means for every vertex~$w \in Q_\Delta$, its adjacency was flipped towards exactly one of~$u$ and~$v$ from~$G$ to~$H$. 
        By construction,~$Q_\Delta$ contains at least two vertices from~$S$, so at least one vertex~$w \in S \setminus \{u,v\} = S \setminus \{v\}$.
        As~$w$ is non-adjacent to both~$u$ and~$v$ in~$G$, we have that~$w$ is adjacent to exactly one of~$u$ and~$v$ in~$H$, so~$u$ and~$v$ are not twins in~$H$.
    \end{claimproof}

    \begin{claimproof}[Proof of \cref{clm:edges}]
        The previous claim also yields a formula~$\sigma(x):=\neg\delta(x)$ that defines the set~$S$ in~$H$. Consider the formula 
        \[
            \pi(x,y) := \forall z : \big[(z \neq x) \wedge (z \neq y) \wedge \sigma(z)\big] \rightarrow \big[E(x,z) \leftrightarrow E(y,z)\big]   
        \]
        expressing that~$x$ and~$y$ have the same neighborhood on~$S \setminus \{x,y\}$.
        We claim that for all~$u \in X$ and~$v \in S$, we have~$H \models \pi(u,v)$ if and only if~$u$ and~$v$ are in the same part of~$\QQ$.
        If~$u$ and~$v$ are in the same part of~$\QQ$, we can verify that~$H \models \pi(u,v)$ using \cref{obs:twin-flip} on the graphs~$H[S \cup \{u,v\}]$ and~$G[S \cup \{u,v\}]$.
        If~$u$ and~$v$ are in different parts, we can verify that~$H \not\models \pi(u,v)$ using \cref{lem:discerning-part} as in the proof of the previous claim.
        
        Next, consider the formula 
        \[
            \phi_\oplus(x,y) := \exists z: \sigma(z) \wedge \pi(z,y) \wedge E(x,z)
        \]
        asking whether there exists a vertex~$z\in S$ that is in the same part of~$\QQ$ as~$y$ and which is adjacent to~$x$.
        We claim that~$\phi_\oplus$ detects between which vertices of~$X$ the adjacency was flipped from~$G$ to~$H$.
        More precisely, for all distinct vertices~$u,v \in X$ we want 
        \[
            H \models \phi_\oplus (x,y) \Leftrightarrow \big(\QQ(u), \QQ(v)\big) \in F.
        \]
        To prove this property, let~$u,v \in X$ be distinct vertices.
        Assume first~$H\models \phi_\oplus(u,v)$. Then there exists a vertex~$z\in S \cap \QQ(v)$ which is adjacent to~$u$ in~$H$.
        As all the vertices of~$S$ are isolated in~$G$, this means the adjacency between~$u$ and~$z$ must have been created by the flip, so we have
       ~$\big(\QQ(u), \QQ(z) = \QQ(v)\big) \in F$.
        Assume now~$\big(\QQ(u), \QQ(v)\big) \in F$.
        By construction there exists a vertex~$z \in \QQ(v) \cap S$.
        As~$z \in S$ is isolated in~$G$, it must be connected to~$u$ in~$H$.
        Then~$z$ witnesses~$H \models \phi_\oplus(u,v)$.
        This proves that~$\phi_\oplus$ has the desired properties.
        We can finish the proof of the claim by setting~$\phi(x,y) := E(x,y) \text{ XOR } \phi_\oplus(x,y)$.
    \end{claimproof}
    Combining the two claims yields the desired interpretation~$I$ satisfying~$I(H) = P_t$.
    We again stress that the construction of~$\delta$ and~$\phi$ was independent of~$t$ (but requires~$t\geq 4$).

    For the ``moreover'' part of the statement, note that, as already observed in the proof of \cref{clm:domain}, all vertices in~$S$ have a twin in~$H$.
    Since~$t\geq 4$,~$S$ has size at least~$8$.
    This finishes the proof of the lemma.
\end{proof}

\subsection{Interpreting Paths in Flipped Half-Graphs}

\begin{definition}
    We call a flipped~$H_t$ \emph{clean} if the adjacency between~$A := \{a_1,\ldots,a_t\}$ and~$B := \{b_1,\ldots,b_t\}$ was not flipped.
    More precisely a graph~$G$ is a clean flipped~$H_t$ if~$G = H_t \oplus (\{A,B\},F)$ for some~$F\subseteq \{(A,A),(B,B)\}$.
\end{definition}

\begin{lemma}\label{lem:clean-flipped}
    Every flipped~$H_{t+1}$ contains an induced clean flipped~$H_{t}$.
\end{lemma}
\begin{proof}
    Let~$G$ be a flipped~$H_{t+1}$.
    If~$G$ is itself clean, then the statement is trivial.
    Otherwise, the subgraph induced by~$\{a_2,\ldots,a_{t+1}, b_1,\ldots,b_{t}\}$ is a clean flipped~$H_{t}$.
\end{proof}

\begin{lemma}\label{lem:paths-from-hg}
    There exists an FO-interpretation~$I$ such that for every~$t\in \N$, every flipped~$H_{t+4}$ contains an induced subgraph~$G$ such that~$I(G)= P_t$.
    
    \smallskip\noindent
    Moreover,~$G$ contains exactly~$4$ vertices that have a twin in~$G$.
\end{lemma}

\begin{figure}[htbp]
    \centering
    \includegraphics{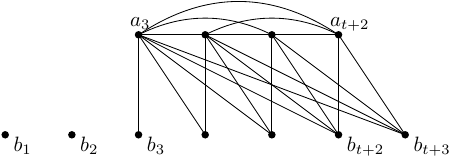}
    \caption{The graph~$G$ from the proof of \cref{lem:paths-from-hg} for~$t=4$.
    It is an induced subgraph of a flipped~$H_{t+3}$. In this example, the set~$A$ is flipped with itself, which is the reason why there is a clique on the~$A$ vertices.}
    \label{fig:interp-from-hg}
\end{figure}

\begin{proof}
    By \cref{lem:clean-flipped}, we can work with a clean flipped~$H_{t+3}$.
    We will interpret~$P_t$ in the subgraph~$G$ induced by the sets~$A:=\{a_3,\ldots,a_{t+2}\}$ and~$B:=\{b_1,\ldots,b_{t + 3}\}$.
    See \cref{fig:interp-from-hg}.
    Let~$\delta(x) := \delta_1(x) \wedge \delta_2(x)$, where
    \begin{gather*}
        \delta_1(x) := \exists y_1 \exists y_2: (y_1 \neq y_2) \wedge \twins(y_1,y_2) \wedge \neg E(y_1,x) \wedge \neg E(y_2,x)\\
        \delta_2(x) := \exists y_3 \exists y_4: (y_3 \neq y_4) \wedge \twins(y_1,y_2) \wedge E(y_3,x) \wedge E(y_4,x) 
    \end{gather*}
    be the formula expressing that there exist a pair of twins that is non-adjacent to~$x$ and a pair of twins that is adjacent to~$x$.
    It is easy to see that this formula defines the set~$A$ in~$G$, i.e., for every~$v\in V(G)$ we have~$G\models \delta(x)$ if and only if~$x \in A$.
    In particular, every satisfying quantification will assign~$b_1$ and~$b_2$ to~$y_1$ and~$y_2$ and~$b_{t+2}$ and~$b_{t+3}$ to~$y_3$ and~$y_4$, which already proves the ``moreover'' part of the statement of the lemma:~$G$ contains exactly~$4$ vertices that have a twin.
    The set~$A$ can be ordered using the formula
    \[
        \sigma(x,y) := x \neq y \wedge \forall z: E(y,z) \rightarrow E(x,z).
    \]
    This means for all~$a_i,a_j \in A$ we have~$G \models \sigma(a_i,a_j)$ if and only if~$i < j$.
    Let~$\phi(x,y) := \mathrm{succ}_\sigma(x,y) \vee \mathrm{succ}_\sigma(y,x)$ be the symmetric closure of the successor of~$\sigma$ expressed by
    \[
        \mathrm{succ}_\sigma(x,y) := \sigma(y,x) \wedge \neg \exists z: \delta(z) \wedge \sigma(y,z) \wedge \sigma(z,x).
    \]
    It is now easy to see that the interpretation defined by~$\delta$ and~$\phi$ satisfies~$I(G)= P_t$.
    As, the construction of~$\delta$ and~$\phi$ was independent of~$t$,~$I$ is the desired interpretation from the statement of the lemma.
\end{proof}

\subsection{Wrapping Up the Interpretation}

We are now ready to prove \cref{prop:patternsToPaths}, which we restate for convenience.

\propPatternsToPaths*

\begin{proof}
    By the pigeonhole principle and hereditariness,~$\CC$ contains either a flipped~$H_t$ for every~$t\in \N$ or a flipped~$tP_t$ for every~$t\in\N$.
    Our interpretation~$I$ works as follows.
    If the input graph~$G$ contains at most~$3$ vertices, then the interpretation leaves~$G$ unchanged.
    Otherwise, it checks whether~$G$ contains exactly~$4$ distinct vertices that have a twin. If this is the case, the interpretation from \cref{lem:paths-from-hg} is used.
    Otherwise, we use the interpretation from \cref{lem:paths-from-swimlanes}.
    It is easily verified that all the conditions are expressible in first-order logic.
    Let us now prove that~$I(\CC)$ contains all paths.
    If~$\CC$ contains arbitrarily large flipped~$tP_t$, then~$I(\CC)$ contains the path of length at most three by \cref{lem:p3-from-swimlane} and every path of length at least four by \cref{lem:paths-from-swimlanes}.
    Otherwise, if~$\CC$ contains arbitrarily large flipped~$H_t$, then~$I(\CC)$ contains the path of length at most three by \cref{lem:p3-from-swimlane} and every path of length at least four by \cref{lem:paths-from-hg}.
\end{proof}

\subsection{Transducing Paths in Flipped \texorpdfstring{$3P_t$s}{3Pts}}

In this subsection we will sharpen our induced subgraph characterization and show the following.
\begin{proposition}
    \label{prop:3ptUnstable}
        Every graph class~$\CC$, that for every~$t\in\N$ contains as induced subgraphs either a flipped~$H_t$ or a flipped~$3P_t$, is monadically MSO-unstable.
\end{proposition}
Note that we do not need to assume that~$\CC$ is hereditary, since we can simulate taking induced subgraphs by colors in the monadic setting.

We have already seen how to~$1$-dimensionally FO-interpret all paths in hereditary classes containing arbitrarily large flipped half-graphs (\cref{lem:p3-from-swimlane,lem:paths-from-hg}). By simulating taking induced subgraphs using colors, this gives us an FO-transduction which producing all paths from every (not necessarily hereditary) class containing arbitrarily large flipped half-graphs. 
Together with \cref{lem:unstable-trans}, this reduces proving \cref{prop:3ptUnstable} to the following.

\begin{lemma}
\label{prop:3ptToPaths}
There is an FO-transduction~$T$ such that for every graph~$G$ that contains an induced flipped~$3P_t$, we have~$P_t \in T(G)$.
\end{lemma}

\begin{proof}
    Let~$G$ be the graph containing an induced subgraph~$H$ that is a flipped~$3P_t$.
    The transduction first marks the three paths of the flipped~$3P_t$ with three colors~$C_1$,$C_2$, and~$C_3$.
    We show how to interpret~$P_t$ in this coloring.
    As the domain we take all the vertices with color~$C_1$.
    Let~$\PP$ be the irreducible~$(H,3P_t)$-flip-partition.
    The formula~$\pi(x,y)$ asking whether~$x$ and~$y$ have the same neighborhood on~$C_3$ checks whether vertices~$x \in C_1$ and~$y\in C_2$ are in the same part of~$\PP$.
    We can now reverse the flips between any two vertices~$x$ and~$y$ on the~$C_1$ path, by checking whether~$x$ is adjacent to the~$C_2$ vertex that is in the same part of~$\PP$ as~$y$.
\end{proof}

\section{Monadic Stability}\label{sec:mstable}
The goal of this section is to show the following.

\begin{proposition}\label{prop:sd-implies-mcmso-stable}
    Every class of bounded shrub-depth is monadically CMSO-stable.
\end{proposition}

It was already shown in \cite{shrubdepth-journal} that no class of bounded shrub-depth CMSO-transduces the class of all paths.
This reduces proving \cref{prop:sd-implies-mcmso-stable} to the following.

\begin{lemma}\label{prop:cmso-transductions}
    Every monadically CMSO-unstable class CMSO-transduces the class of all paths.
\end{lemma}

For every CMSO-unstable class~$\CC$, there is a CMSO-formula~$\phi(\bar x, \bar y)$ that orders arbitrarily large sequences of tuples in a coloring of~$\CC$.
This is already close to what we want: the successor relation definable through~$\phi$ forms a ``path'' on these tuples.
However, we cannot directly produce this path by a transduction, as transductions only work with singletons and not with tuples.
The trick is to use the color predicates that are available in the monadic setting, to show the existence of a formula~$\phi'(x,y)$ with single free variables that has the order-property on another coloring of~$\CC$.
For FO-formulas, it was shown that this is possible by Baldwin and Shelah \cite[Lem.\ 8.1.3]{baldwin1985second} (see also \mbox{\cite[Thm.\ 2.2]{anderson1990tree}}), and Simon proved a strengthening of the statement that involves parameters instead of color predicates~\cite{simon2021note}.
Both proofs work in a model theoretic setting on infinite structures.
However, their core is combinatorial and generalizes from FO to more expressive logics without problems, which yields the following lemma.

\begin{restatable}{lemma}{lemMstableTransductions}
\label{lem:mstable-transductions}
    For every logic~$\LL$ that extends FO and every class of~$\Sigma$-structures~$\CC$,~$\CC$ is monadically~$\LL$-stable if and only if~$\CC$ does not~$\LL$-transduce the class of all half-graphs.
\end{restatable}

The proof of \cref{lem:mstable-transductions} is mostly a translation of Simon's result~\cite{simon2021note} from the infinite to the finite setting and can be found in the appendix.

In \cref{lem:mstable-transductions}, we say a logic~$\LL$ \emph{extends FO} if for every relational symbol~$R(\bar x)$,~$\FO[\Sigma \cup \{ R \}]$-formula~$\phi(\bar y)$, and~$\LL[\Sigma]$-formula~$\rho(\bar x)$, the formula~$\phi'(\bar y)$ obtained by replacing each occurrence of the relation~$R(\bar x)$ in~$\phi$ with~$\rho(\bar x)$ is a~$\LL[\Sigma]$-formula.
This definition is a bit fuzzy, as we do not define what it means to be a \emph{logic}, but the reader will agree that FO, MSO, and CMSO all extend FO. The same holds for other natural extensions of FO, such as separator logic \cite{schirrmacher23,bojanczyk2021separator}.

\cref{lem:mstable-transductions} implies \cref{prop:cmso-transductions} as follows: every monadically CMSO-unstable graph class~$\CC$ CMSO-transduces the class of all half-graphs~$\mathcal{H}$.
Clearly,~$\mathcal{H}$ CMSO-transduces the class of all paths~$\PP$. Using \cref{lem:interp-mso}, we obtain a CMSO-transduction from~$\CC$ to~$\PP$.

\paragraph*{An Alternative Proof of \cref{prop:sd-implies-mcmso-stable}.}
Let us briefly mention a second way of proving \cref{prop:sd-implies-mcmso-stable}, by showing that~$\infty$-flip-flat classes are monadically CMSO-stable.
Through iteration, we can prove a variant of~$\infty$-flip-flatness for tuples.
Then we can use Feferman-Vaught style theorems~\cite{MAKOWSKY2004159} to show that for a fixed~$k\in\N$, no CMSO-formula can order arbitrarily large~$(\infty,k)$-flip-flat sequence of tuples. 
This mirrors the use of Gaifman's locality theorem~\cite{gaifman82} in the setting of monadically FO-stable classes~\cite{dreier2022indiscernibles}.

\paragraph*{Monadic Dependence.}
\emph{Dependence} is another model theoretic dividing line, which generalizes stability.
A formula~$\phi(\bar x, \bar y)$ has the \emph{$\ell$-independence-property} on a structure~$G$, if there exist elements~$\bar a_i \in V(G)^{|\bar x|}$ for each~$i \in [\ell]$ and~$\bar b_S \in V(G)^{|\bar y|}$ for each~$S \subseteq [\ell]$ such that for~$i \in [\ell], S \subseteq [\ell]$
\[
    G \models \phi(\bar a_i, \bar b_S) \Leftrightarrow i \in S.
\]
Similarly,~$\phi$ has the \emph{independence-property} on a class of structures~$\CC$, if for every~$\ell \in \N$, there exists a structure in~$\CC$ on which~$\phi$ has the~$\ell$-independence-property.
We say~$\CC$ is \emph{$\LL$-dependent}\footnote{Dependence is also known as \emph{NIP}, which stands for ``Not the Independence Property''.} 
if no~$\LL$-formula has the independence-property on~$\CC$.
The definition of a \emph{monadically}~$\LL$-independent class is as expected.
Simon proves his result not only for stability, but also for dependence~\cite{simon2021note}.
The translation to the finite that we do in the appendix yields the following analog of \cref{lem:mstable-transductions} for dependence.

\begin{restatable}{lemma}{lemMdepTransductions}
    \label{lem:mdep-transductions}
    For every logic~$\LL$ that extends FO and every class of~$\Sigma$-structures~$\CC$,~$\CC$ is monadically~$\LL$-dependent if and only if~$\CC$ does not~$\LL$-transduce the class of all graphs.
\end{restatable}

FO-dependence has recently gained traction in structural graph theory~\cite{dreier2024flipbreakability,twwIV,bonnet2022model}, and
we believe that \cref{lem:mdep-transductions} will be useful for the future study of (C)MSO-dependence, which we comment on in the outlook section of this paper.

\section{The Expressive Power of MSO}\label{sec:expressiveness}

The goal of this section is to prove the following.

\begin{proposition}\label{prop:sep-fo-mso}
    MSO is more expressive than FO on every hereditary graph class, that for every~$t\in\N$ contains either a flipped~$H_t$ or a flipped~$tP_t$, 
\end{proposition}

By the pigeonhole principle and hereditariness, it is sufficient to prove
the above lemma in two separate cases:
either the graph class contains arbitrarily large flipped~$H_t$ or it contains arbitrarily large flipped~$tP_t$.
We will do so in the upcoming \cref{lem:sep-hg,lem:sep-tpt}.
For this purpose we first show how interpretations can be used to prove inexpressibility results.

\subsection{Interpretations and Inexpressibility}

\begin{lemma}\label{lem:interp-type}
    For every FO-interpretation~$I$ from~$\Sigma_1$ to~$\Sigma_2$ and~$q \in \N$ there exists~$q' \in \N$ such that for every two~$\Sigma_1$-structures
   ~$G$ and~$H$
    \[
        \tp(G,\FO_{q'}) = \tp(H,\FO_{q'}) \Rightarrow 
        \tp(I(G),\FO_{q}) = \tp(I(H),\FO_{q}). 
    \]
\end{lemma}
\begin{proof}
    Consider the set~$\Phi$ of~$\FO[\Sigma_2]_q$-sentences.
    Up to equivalence, we can assume~$\Phi$ is finite.
    Let~$\Phi' := \{\phi_I : \phi \in \Phi\}$ where~$\phi_I$ is obtained from \cref{lem:interp-fo}.
    Choose~$q'$ to be the largest quantifier rank among the sentences in~$\Phi'$.
    \cref{lem:interp-fo} now gives
    \[
        \tp(I(G),\FO_q) \neq \tp(I(H),\FO_q) \Rightarrow 
        \tp(G,\FO_{q'}) \neq \tp(H,\FO_{q'}),
    \]
    and the lemma holds by contraposition.
\end{proof}

Denote by~$\struc L_t$ the \emph{linear order of length~$t$} represented as the structure with universe~$[t]$ and a single binary relation~$<$ interpreted as expected.
FO cannot distinguish long linear orders:

\begin{theorem}[{\cite[Thm.\ 3.6]{libkin2004elements}}]
    For every~$q \geq 0$ and~$s,t \geq 2^q$, we have~$\tp(\struc L_{s}, \FO_q) = \tp(\struc L_{t}, \FO_q)$.
\end{theorem}

\begin{lemma}\label{lem:interp-order-to-path}
    There exists a~$1$-dimensional FO-interpretation~$I$ with~$I(\struc L_t) = P_t$ for every~$t\in \N$.
\end{lemma}
\begin{proof}
    We set~$\delta_I(x) := \text{true}$ and~$\phi_{I,E} := \mathrm{pred}_<(x,y) \vee \mathrm{pred}_<(y,x)$ where 
    \[
        \mathrm{pred}_<(x,y) := x < y \wedge \neg \exists z: x < z \wedge z < y.\qedhere
    \]
\end{proof}

\newcommand{\even}{\phi_{\mathrm{even}}}
\newcommand{\sameparity}{\phi_{\mathrm{sameParity}}}

\begin{lemma}\label{lem:even}
    There is an MSO-sentence~$\even$ such that for every~$t \in \N$,~$P_t\models \even$ if and only if~$t$ is even.
\end{lemma}
\begin{proof}
    The sentence~$\even$ existentially quantifies a partition of the vertex set into two parts, demands that no part contains two adjacent vertices, and that the endpoints of the path (definable as the two only vertices with degree~$1$) are in different parts.
\end{proof}

\begin{corollary}\label{lem:same-parity}
    There is an MSO-sentence~$\sameparity$ such that for every graph~$G$ that is a disjoint union of paths,~$G \models \sameparity$ if and only if
    \begin{itemize}
        \item all connected components of~$G$ are paths of even length, or 
        \item all connected components of~$G$ are paths of odd length.
    \end{itemize}
\end{corollary}
\begin{proof}
    We first construct a sentence~$\phi_{\mathrm{allEven}}$ that checks whether all components of~$G$ have even length.
    The sentence demands that for every vertex~$v\in V(G)$,~$\even$ holds true when relativized to the connected component of~$v$.
    Here we express connectivity between two vertices~$x,y$ by a formula demanding that every set that contains~$x$ and is closed under taking neighbors must also contain~$y$.
    We can similarly express~$\phi_{\mathrm{allOdd}}$ and set~$\sameparity := \phi_{\mathrm{allEven}} \vee \phi_{\mathrm{allOdd}}$.
\end{proof}

As an example, we will now show how to use interpretations to separate MSO from FO on the class of all paths.

\begin{lemma}\label{lem:expressiveness-paths}
    MSO is more expressive than FO on the class of all paths.
\end{lemma}
\begin{proof}
    Assume towards a contradiction the existence of an FO-sentence~$\psi$ that is equivalent to the MSO-sentence~$\even$ from \cref{lem:even} on class of all paths, and 
    let~$q$ be the quantifier rank of~$\psi$.
    Let~$I$ be the interpretation from the class of all linear orders to the class of all paths from \cref{lem:interp-order-to-path} and let~$q'$ be the quantifier-rank obtained from \cref{lem:interp-type} to~$I$ and~$q$.
    Let~$s:=2^{q'}$ and~$t := s + 1$.
    By \cref{lem:interp-order-to-path}, we have that~$\tp(\struc L_s, \FO_{q'}) = \tp(\struc L_t, \FO_{q'})$ and, by \cref{lem:interp-type}, also 
    \[
        \tp(I(\struc L_s) = P_s, \FO_{q}) = \tp(I(\struc L_t) = P_t, \FO_{q}).    
    \]
    In particular~$\psi$ does not distinguish~$P_s$ and~$P_t$.
    However,~$\even$ distinguishes~$P_s$ and~$P_t$, as exactly one of~$\{s,t\}$ is even.
    Hence,~$\even$ and~$\psi$ are not equivalent on the class of all paths; a contradiction.
\end{proof}

\subsection{Separating FO and MSO on Flipped Half-Graphs}

\begin{restatable}{lemma}{lemSepHg}
    \label{lem:sep-hg}
    Let~$\CC$ be a hereditary graph class that contains a flipped~$H_t$ for every~$t\in\N$.
    Then MSO is more expressive than FO on~$\CC$.
\end{restatable}

\begin{figure}[htbp]
    \centering
    \includegraphics{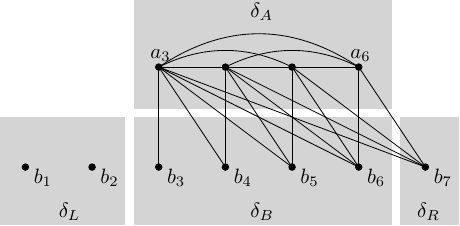}
    \caption{$H^\star_4$ together with a partitioning of its vertex set by the domain formulas.}
    \label{fig:hgs-from-lo}
\end{figure}

\begin{proof}
    By \cref{lem:clean-flipped}, we can assume that~$\CC$ contains a clean flipped~$H_t$ for every~$t \in \N$.
    Recall that in a clean flipped~$H_t$, the adjacencies between~$A := \{a_1,\ldots,a_t\}$ and~$B := \{b_1,\ldots,b_t\}$ are not flipped.
    This means, for each~$t\in\N$ there exist four different clean flipped~$H_t$, each of the form~$H_t \oplus (\{A,B\},F)$ for one of the four possible subsets~$F \subseteq \{(A,A),(B,B)\}$.
    We call~$F$ the \emph{flavor} of a clean flipped~$H_t$.
    By the pigeonhole principle and hereditariness, we can assume that there is a single flavor~$F$ such that~$\CC$ contains the clean flipped~$H_t$ of flavor~$F$ for every~$t\in\N$.
    We will only show the proof for~$F = \{(A,A)\}$, but it is easily adapted to the other three flavors.

    For every~$t \in \N$, we denote by~$H_t^\star$ the subgraph of the
    clean flipped~$H_{t+3}$ of flavor~$F$ induced by the sets~$\{a_3,\ldots,a_{t+2}\}$ and~$\{b_1,\ldots,b_{t + 3}\}$.
    See \cref{fig:hgs-from-lo} for an illustration.
    We have already shown in \cref{lem:paths-from-hg} the existence of a~$1$-dimensional FO-interpretation~$I^\star$ such that for all~$t\in\N$ we have~$I^\star(H^\star_t) = P_t$.
    Combining this insight with \cref{lem:even} and \cref{lem:interp-mso} yields an MSO-sentence~$\even^\star$ such that for every~$t \in \N$ we have~$H^\star_t \models \even^\star$ if and only if~$t$ is even.
    We claim that no FO-sentence is equivalent to~$\even^\star$ on~$\CC$.
    In order to continue the proof in the same way as in the proof of \cref{lem:expressiveness-paths}, it suffices to show the existence of an FO-interpretation~$I$ such that for every~$t \geq 2$ we have~$I(\struc L_t) = H^\star_t$.
    We will construct~$I$ as a~$5$-dimensional interpretation.
    As~$\struc L_t$ has only~$t$ elements and~$H^\star_t$ has~$2t + 3$ elements, we will use the~$5$ dimensions to encode additional elements as follows.
    For~$\bar s \in \{*,0,1\}^5$ let 
    \[
        \delta_{\bar s}(\bar x) := \bigwedge_{i \in [5]} 
        \big (\bar s[i] = 0 \rightarrow \min(\bar x[i])\big )
        \wedge 
        \big(\bar s[i] = 1 \rightarrow \max(\bar x[i])
        \big ) 
    \]
    where~$\min(x)$ and~$\max(x)$ are replaced with the easily definable FO-formulas checking whether~$x$ is the smallest or largest element of the linear order, respectively.
    This means the formula~$\delta_{\bar s}(\bar x)$ checks which elements of~$\bar x$ are equal to the minimum/maximum element, as specified in~$\bar s$, where~$*$ is used as a wildcard.
    The two formulas~$\delta_A := \delta_{*0000}$ and~$\delta_B := \delta_{*1111}$ are each satisfied by~$t$ many tuples,~$\delta_L := \delta_{10100} \vee \delta_{10101}$ is satisfied by two tuples, and~$\delta_R := \delta_{10110}$ is only satisfied by a single tuple.
    Moreover, each tuple satisfies at most one of~$\delta_A,\delta_B,\delta_L,\delta_R$.
    This means setting the domain formula of~$I$ to be
    \[
        \delta_I(\bar x) := \delta_A(\bar x) \vee \delta_B(\bar x) \vee
        \delta_L(\bar x) \vee \delta_R(\bar x)
    \]
    creates a universe of the desired size~$2t+3$.
    See \cref{fig:hgs-from-lo} for how the elements of the domain will map to the elements of~$H^\star_t$.
    We set the edge formula of the interpretation to be~$\phi_{I,E}(\bar x, \bar y) := \phi(\bar x, \bar y) \vee \phi(\bar y, \bar x)$, where
    \[
        \phi(\bar x, \bar y) := 
        \big(
            \delta_A(\bar x) \wedge \delta_B(\bar y) \wedge \bar x[1] \leq \bar y[1]
        \big)
        \vee
        \big(
            \delta_A(\bar x) \wedge \delta_A(\bar y)
        \big)
        \vee
        \big(
            \delta_A(\bar x) \wedge \delta_R(\bar y)
        \big),
    \]
    and where~$(x \leq y)$ is a shorthand for~$(x < y \vee x = y)$.
    The formula~$\phi$ can be easily adapted to the different flavors~$F$.
    It is now easy to verify that~$I(\LL_t) = H^\star_t$ for every~$t\geq 2$.

    With the MSO-sentence~$\even^\star$ and the interpretation~$I$ in place, the proof can now be finished in the same way as the proof of \cref{lem:interp-order-to-path}.
\end{proof}

\subsection{Separating FO and MSO on Flipped \texorpdfstring{$tP_t$}{tPt}s}

In this subsection we prove the following.

\begin{restatable}{lemma}{lemSepTpt}
    \label{lem:sep-tpt}
    Let~$\CC$ be a hereditary graph class that contains a flipped~$tP_t$ for every~$t\in\N$.
    Then MSO is more expressive than FO on~$\CC$.
\end{restatable}

We recall the definition of a (flipped)~$mP_t$ as we need to precisely refer to its vertices.
\defSwimlanes*

\newcommand{\nibble}{\mathrm{nibble}}
Let~$G$ be a flipped~$mP_t$ with~$m\geq 2$ and~$t \geq 3$.
For~$i \in \{1,2\}$, we denote by~$\nibble_i(G)$ the graph obtained by removing from~$G$ the vertices~$(1,1)$ and~$(i,t)$.
Crucially, FO cannot distinguish between the two nibbles of a flipped~$mP_t$ for large~$t$, as made precise by the following lemma.

\begin{restatable}{lemma}{lemNibbleEquiv}
    \label{lem:nibble-equiv}
    For every~$q\geq 1$ and every graph~$G$ that is a flipped~$mP_t$ with~$m\geq 2$ and~$t \geq 3^q$
    \[
        \tp(\nibble_1(G),\FO_q) = \tp(\nibble_2(G),\FO_q).
    \]
\end{restatable}

The proof of \cref{lem:nibble-equiv} requires some additional tooling and is deferred to \cref{sec:nibble-equiv}.

\begin{lemma}\label{lem:sep-tpt-sentence}
    There exists an MSO sentence~$\phi$ such that for every~$m \geq 9$,~$t \geq 3$, and graph~$G$ that is a flipped~$mP_t$
    \[
        \nibble_1(G) \models \phi \quad\text{and}\quad \nibble_2(G) \not \models \phi.
    \]
\end{lemma}
\begin{proof}
    Fix~$t$ and~$G$ as in the statement of the lemma. 
    Let~$G_i := \nibble_i(G)$ for~$i\in\{1,2\}$, let~$G^\bullet_i$ be the ``non-flipped version'' of~$G_i$. More precisely~$G^\bullet_i$ is the graph~$mP_t$ with vertices~$(1,1)$ and~$(i,t)$ removed.
    Crucially,~$G^\bullet_1$ is the disjoint union of~$(m-1)P_t$ and~$P_{t-2}$ and~$G^\bullet_2$ is the disjoint union of~$(m-2)P_t$ and~$2P_{t-1}$, and therefore
    \[
        G^\bullet_1 \models \sameparity 
        \quad\text{and}\quad
        G^\bullet_2 \not\models \sameparity,
    \]
    where~$\sameparity$ is the MSO-sentence from \cref{lem:same-parity} that checks whether all connected components of~$G_i^\bullet$ have the same parity.
    By \cref{lem:interp-mso}, it now suffices to give a~$1$-dimensional MSO-interpretation~$I$ such that~$I(G_i) = G_i^\bullet$.
    We next construct such an~$I$, which is even an FO-interpretation.

    Let~$\PP_i$ and~$F_i$ be the irreducible~$(G_i,G^\bullet_i)$-flip-witnesses
    and let~$\pi(x,y)$ be the formula expressing that the neighborhoods of~$x$ and~$y$ disagree on at most~$4$ vertices in~$V(G_i) \setminus \{x,y\}$.
    \begin{claim}
        For all~$u,v \in V(G_i)$,~$G_i \models \pi(u,v)$ if and only if~$u,v$ are in the same part of~$\PP_i$.
    \end{claim}
    \begin{claimproof}
        If~$u = (p,q)$ and~$v = (p',q')$ are in the same part of~$\PP_i$ then the only vertices where they can disagree are the four vertices~$(p,q-1)$,~$(p,q+1)$,~$(p',q'-1)$, and~$(p',q'+1)$, i.e., their neighbors in~$G^\bullet_i$.
        If~$u$ and~$v$ are in different parts~$Q_1$ and~$Q_2$ of~$\PP_i$ then there exists a discerning part~$Q_\Delta$ by
        \cref{lem:discerning-part}.
        Due to the structure of a flipped~$mP_t$, the part
       ~$Q_\Delta$ contains at least one layer~$L_\ell$.
        Then~$u$ and~$v$ must disagree  on the at least~$9-4=5$ vertices 
       ~$L_\ell \setminus \big(\{1,2,p,p'\} \times [\ell]\big) \subseteq V(G_i) \setminus \{ u,v \}$.
    \end{claimproof}
    Using~$\pi(x,y)$, we can construct a formula~$\phi_\oplus(x,y)$ checking whether~$x$ has at least three neighbors in~$\PP_i(y)$.
    We claim that this formula detects, whether the connection between two vertices was flipped.
    \begin{claim}
        For all~$u,v \in V(G_i)$,~$G_i \models \phi_\oplus(u,v)$ if and only if~$(\PP_i(u),\PP_i(v)) \in F_i$.
    \end{claim}
    \begin{claimproof}
        If the adjacency between~$u = (p,q)$ and~$v = (p',q')$ was flipped then~$u$ is adjacent in~$G_i$ to all the at 
        least~$5$ vertices
       ~$L_{q'} \setminus \big(\{1,2,p,p'\} \times [q']\big)\subseteq \PP_i(v) \setminus \{ u,v \} \subseteq V(G_i) \setminus \{ u,v \}$.
        If the adjacency between~$u$ and~$v$ was not flipped then the adjacency between~$u$ and~$\PP_i(v)$ is the same as in~$G_i^\bullet$, where~$u$ has at most two neighbors.
    \end{claimproof}
    We now define the~$1$-dimensional FO-interpretation~$I$ by setting 
    \[
        \delta_I(x) := \mathrm{true} \quad\text{and}\quad \phi_{I,E} := E(x,y) \XOR \phi_\oplus(x,y).    
    \]
    Using the above claim it is easy to check that~$I(G_i) = G_i^\bullet$, as desired.
    Applying \cref{lem:interp-mso} to~$I$ and~$\sameparity$, yields a sentence~$\phi$ such that~$G_1 \models \phi$ but~$G_2 \not \models \phi$.
    This proves the lemma, as the definition of~$I$ (and therefore also the definition of~$\phi$) depends on neither of~$m,t,G$.
\end{proof}

We can now prove \cref{lem:sep-tpt}.
\lemSepTpt*
\begin{proof}
    Assume towards a contradiction the existence of an FO-sentence~$\psi$ equivalent on~$\CC$ to the MSO-sentence~$\phi$ from \cref{lem:sep-tpt-sentence}. 
    Let~$q$ be the quantifier rank of~$\psi$,~$t := 3^q$,
    and~$G$ be a flipped~$9P_t$ that is contained in~$\CC$ by the assumption of the lemma.
    By hereditariness, we have that~$G_i := \nibble_i(G)$ is contained in~$\CC$ for~$i\in\{1,2\}$.
    By \cref{lem:sep-tpt-sentence},~$\phi$ distinguishes between~$G_1$ and~$G_2$, but by \cref{lem:nibble-equiv},~$\psi$ does not; a contradiction.
\end{proof}

\subsection{Proof of \texorpdfstring{\cref{lem:nibble-equiv}}{lem:nibble-equiv}}\label{sec:nibble-equiv}
An \emph{isomorphism type} for a signature~$\Sigma$ is an equivalence class for the ``is isomorphic to'' relation on~$\Sigma$-structures, i.e., two~$\Sigma$-structures have the same isomorphism type if and only if they are isomorphic.
The signature~$\Sigma$ will often be clear from the context and omitted.

For a~$k$-colored graph~$G$,~$r\in \N$,~$v\in V(G)$, the \emph{$r$-ball marked at~$v$ in~$G$} is the substructure induced by the radius-$r$ neighborhood~$N_r[v]$ in~$G$, where~$v$ is marked as a constant.
This means the~$r$-ball marked at~$v$ is a structure over the signature~$\Gamma^{(k)} \cup \{c\}$ with an additional constant symbol~$c$.
For an isomorphism type~$\tau$, a~$k$-colored graph~$G$, and~$r\in\N$, we write~$\#(\tau,r,G)$ to denote the number of vertices~$v$ in~$G$ such that the~$r$-ball marked at~$v$ in~$G$ has isomorphism type~$\tau$.

The following theorem was proven by Fagin, Stockmeyer, and Vardi \cite[Thm.\ 4.3]{FAGIN199578} in the more general setting of arbitrary finite structures (and not just colored graphs).
It is based on the ideas of Hanf \cite{HANF2014132} for infinite structures.
See also \cite[Thm.\ 4.24]{libkin2004elements}.


\begin{theorem}[Finite Hanf locality]
    For every~$q,d \in \N$ there exists~$m\in\N$ such that for every~$k\in\N$,~$k$-colored graphs~$G_1$ and~$G_2$ with maximum degree~$d$ if 
    \[
        \#(\tau,3^{q-1},G_1) = \#(\tau,3^{q-1},G_2)
        \quad \text{ or } \quad
        \big(
            \#(\tau,3^{q-1},G_1)\geq m  \text{ and }\#(\tau,3^{q-1},G_2) \geq m    
        \big)
    \]
    for every isomorphism type~$\tau$, then 
   ~$\tp(G_1,\FO[\Gamma^{(k)}]_q) = \tp(G_2,\FO[\Gamma^{(k)}]_q)$.
\end{theorem}

We will only use the theorem in cases where we can guarantee that the number of isomorphism types is exactly the same.
We can therefore use the following simplified version of the theorem, obtained by setting~$d$ to be the maximum of the maximum degrees of~$G_1$ and~$G_2$.

\begin{corollary}\label{cor:hanf}
    For every~$q,k\in \N$ and every two~$k$-colored graphs~$G_1$ and~$G_2$,
    if~$\#(\tau,3^{q-1},G_1) = \#(\tau,3^{q-1},G_2)$ for every isomorphism type~$\tau$, then~$\tp(G_1,\FO[\Gamma^{(k)}]_q) = \tp(G_2,\FO[\Gamma^{(k)}]_q)$.
\end{corollary}

We have gathered all the ingredients to prove \cref{lem:nibble-equiv}, which we restate for convenience.

\lemNibbleEquiv*

\begin{proof}
    Let~$q,G,m,t$ be as in the statement
    For~$i\in \{1,2\}$ we set~$G_i := \nibble_i(G)$ and define~$G^+_i$ to be the~$t$-colored graph obtained from~$G_i$ by reversing all the flips made to obtain~$G$ from~$mP_t$, and marking the~$t$ layers of~$mP_t$ each with an individual color.
    This means~$G_1^+$ is a~$t$-coloring of the disjoint union of~$m - 1$ many~$t$-vertex paths and one~$(t-2)$-vertex path, and~$G_2^+$ is a~$t$-coloring of the disjoint union of~$m - 2$ many~$t$-vertex paths and two~$(t-1)$-vertex paths.
    Let us argue that in order to prove the lemma, it is sufficient to show 
    \begin{equation}\label{eq:nibble-equiv}
        \tag{$*$}
        \tp(G_1^+,\FO[\Gamma^{(t)}]_q) = \tp(G_2^+,\FO[\Gamma^{(t)}]_q).    
    \end{equation}
    This is because each~$\FO_q$-sentence~$\phi$ over the signature of graphs, can be rewritten into an
   ~$\FO[\Gamma^{(t)}]_q$-sentence~$\phi^+$ such that for both~$i\in\{1,2\}$ we have
        $G_i \models \phi \Leftrightarrow G_i^+ \models \phi^+$.
    We obtain~$\phi^+$ by replacing each occurrence of~$E(x,y)$ in~$\phi$ with 
    $\big(E(x,y) \XOR \phi_\oplus(x,y)\big)$,
    where~$\phi_\oplus(x,y)$ is the formula that determines from the vertex-colors of~$x$ and~$y$ in~$G^+_i$ whether the adjacency between the layers containing~$x$ and~$y$ was flipped going from~$mP_t$ to~$G$.
    Note that the length of the formula~$\phi_\oplus(x,y)$, as well as the size of the signature~$\Gamma^{(t)}$, depends on the size of input graph~$G$. This is unusual, but it will not cause us any problems as we do not introduce new quantifiers:~$\phi^+$ has the same quantifier rank as~$\phi$.

    \begin{figure}[h!]
        \centering
        \includegraphics[scale = 1.3]{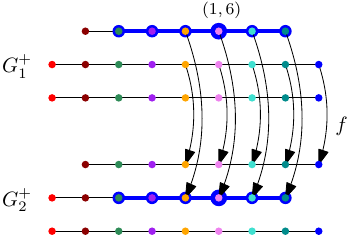}
        \caption{An illustration of the case~$q = 2$,~$m = 3$,~$t = 9 \geq 3^q$.
        Depicted are the~$t$-colored graphs~$G^+_1$ and~$G^+_2$. The bijection~$f : V(G^+_1) \to  V(G^+_2)$ is drawn between the two graphs, where we have omitted the mapping arrows for vertices~$(i,j)$ from~$G^+_1$ that are mapped to the same coordinates~$(i,j)$ in~$G^+_2$.
        We have highlighted the~$(3^{q-1} = 3)$-ball marked at the vertex~$(1,6)$ in~$G_1^+$ as well as the isomorphic~$3$-ball marked at~$f((1,6))$ in~$G^+_2$.
        }
        \label{fig:nibble-iso}
    \end{figure}

    In order to prove \eqref{eq:nibble-equiv}, by \cref{cor:hanf}, it suffices to construct a bijection~$f : V(G_1^+) \to V(G_2^+)$ such that for every vertex~$v\in V(G_1^+)$, the~$3^{q-1}$-ball marked at~$v$ in~$G_1^+$ has the same isomorphism type as the~$3^{q-1}$-ball marked at~$f(v)$ in~$G_2^+$.
    We choose
    \[
        f\big((i,j)\big) := \begin{cases}
            (2,j) & \text{if~$i=1$ and~$j \geq 3^q / 2$,}\\
            (1,j) & \text{if~$i=2$ and~$j \geq 3^q / 2$,}\\
            (i,j) & \text{otherwise.}
        \end{cases}
    \]
    The bijection maps the right half of the first (second) path in~$G_1^+$ to the right half of the second (first) path in~$G_2^+$, and is
    depicted in \cref{fig:nibble-iso}. 
    It is easy to verify that~$f$ has the desired properties.
    This concludes the proof of the lemma.
\end{proof}

\section{Outlook: Stronger Obstructions and MSO-Dependence}\label{sec:outlook}
In this work, we have characterized graph classes of bounded shrub-depth by forbidden induced subgraphs.
This has allowed us to derive further characterizations through logic, for instance by the model theoretic property of MSO-stability, and by comparing the expressive power of FO and MSO.
While the obstructions we found were sufficient to achieve our goal of characterizing MSO-stability, we initially had stronger obstructions in mind, which we could neither prove nor refute:

\begin{conjecture}
    For every graph class~$\CC$ of unbounded shrub-depth there exists~$k\in\N$ such that~$\CC$ contains as induced subgraphs either
    \begin{itemize}
      \item a flipped~$H_t$ for every~$t\in\N$, or
      \item a~$k$-flip of~$P_t$ for every~$t \in \N$.
    \end{itemize}
\end{conjecture}

As we have seen in \cref{lem:fo-unstable-to-ideal-patterns}, this conjecture holds for every class that is monadically FO-unstable.
Even stronger, the~$k$-flips of~$P_t$s appearing there can be assumed to be \emph{periodic}.
This means, the flip-partition splits the paths in a repeating pattern. This is due to the fact that we have extracted the~$k$-flips of~$P_t$, by ``snaking'' along the known obstructions characterizing monadic FO-stability (see \cref{fig:fo-patterns}).
The advantage of these periodic flips is that they can be finitely described:
for every monadically FO-unstable class~$\CC$ there exists an algorithm that given~$t\in\N$, returns a size-$t$ obstruction (a flipped~$H_t$ or a~$k$-flip of~$P_t$) that is contained in~$\CC$.
We currently do not know whether classes of unbounded shrub-depth admit such algorithms, as we have little control over how the~$tP_t$s are flipped.
Having such a finite description would help to lift algorithmic hardness results from the class of all paths to every hereditary classes of unbounded shrub-depth.
One such result is by Lampis, who shows that there is no fpt MSO model checking algorithm for the class of all paths whose runtime dependence is elementary on the size of the input formula, under the complexity theoretic assumption E~$\neq$ NE~\cite{lampis2014lower}.
Lifting this hardness result to hereditary classes of unbounded shrub-depth would yield another characterization, as it is known that every class of bounded shrub-depth admits elementary fpt MSO model checking~\cite{gajarsky2015kernelizing,shrubdepth}.

A second interesting question concerns the model theoretic notion of \emph{dependence} (see \cref{sec:mstable} for a definition).
Similar to FO-stable classes, also FO-dependent classes have recently been shown to admit nice combinatorial characterizations~\cite{dreier2024flipbreakability}.
This raises the question whether also MSO-dependence can be combinatorially characterized.
It is natural to conjecture the following.

\begin{conjecture}\label{conj:cw}
  For every hereditary graph class~$\CC$, the following are equivalent.
  \begin{enumerate}
    \item $\CC$ has bounded clique-width (or equivalently bounded rank-width).
    \item $\CC$ is MSO-dependent.
    \item $\CC$ is monadically MSO-dependent.
    \item $\CC$ is CMSO-dependent.
    \item $\CC$ is monadically CMSO-dependent.
  \end{enumerate}
\end{conjecture}

It is already known that a class has bounded clique-width if and only if it does not CMSO-transduce the class of all graphs~\cite{COURCELLE200791}.
By \cref{lem:mdep-transductions}, this confirms the equivalence~$(1)\Leftrightarrow(5)$ of the conjecture.
We remark that, again by \cref{lem:mdep-transductions}, both of the two equivalence~$(1)\Leftrightarrow (2)$ and~$(1)\Leftrightarrow (3)$ would imply that every class of unbounded clique-width MSO-transduces the class of all graphs.
In turn, this would imply the longstanding Seese's conjecture stating that the MSO-theory of every graph class of unbounded clique-width is undecidable \cite{seese1991conjecture, courcelle2012graph}.
This suggests that \cref{conj:cw} is a tough nut to crack.

\bibliographystyle{plain}
\bibliography{ref}

\newpage
\appendix

\section{Monadic Stability via Transductions}

\newcommand{\Th}{\mathrm{Th}}

The goal of this section is to prove the following lemma used in \cref{sec:mstable}.

\lemMstableTransductions*

Fix a logic~$\LL$ and a signature~$\Sigma$.
For~$\ell\in \N\cup\{\infty\}$, an~$\LL[\Sigma]$-formula~$\phi(\bar x,\bar y, \bar z)$ has the \emph{bipartite~$\ell$-order-proprety} on a~$\Sigma$-structure~$G$ if there exist tuples~$\bar a_i \in V(G)^{|\bar x|}$,~$\bar b_i \in V(G)^{|\bar y|}$ for every~$i \in [\ell]$, and~$\bar c \in V(G)^{|\bar z|}$, such that for all~$i,j \in [\ell]$
\[
    G \models \phi(\bar a_i, \bar b_j, \bar c) \Leftrightarrow i \leq j.
\]
The formula~$\phi$ has the \emph{bipartite order-property} on a class of~$\Sigma$-structures~$\CC$, if for every~$\ell\in\N$ there is a structure in~$\CC$ on which~$\phi$ has the bipartite~$\ell$-order-property.
Note that whenever a formula~$\phi(\bar x, \bar y)$ has the~$\ell$-order-property witnessed by tuples~$\bar d_i$ on a structure~$G$, then~$\phi$ also has the bipartite~$\ell$-order-property on~$G$ witnessed by the tuples~$\bar a_i = \bar b_i = \bar d_i$ and where the parameters~$\bar c$ are the empty tuple.

A \emph{theory} is a set of FO-sentences.
A structure~$M$ is a \emph{model} of a theory~$T$, if~$M$ satisfies all sentences in~$T$.
The compactness theorem states that a theory~$T$ has a model if and only if every finite subset of~$T$ has a model.
A formula~$\phi(\bar x, \bar z, \bar y)$ has the \emph{bipartite order-property} on a theory~$T$ if there exists a model~$M$ of~$T$ on which~$\phi$ has the bipartite~$\infty$-order-property.

The following theorem is by Simon~\cite{simon2021note}. 
\begin{theorem}[\cite{simon2021note}]\label{thm:simon}
    For every theory~$T$, if there is an FO-formula~$\phi(\bar x, \bar y, \bar z)$ that has the bipartite order-property on~$T$, then there also exists an FO-formula~$\psi(x,y,\bar z')$ with singleton variables~$x$ and~$y$ that has the bipartite order-property on~$T$.
\end{theorem}
We translate it from the infinite setting of theories to classes of finite structures, and extend it to more expressive logics.

\begin{lemma}\label{lem:singletons}
    For every logic~$\LL$ that extends FO, signature~$\Sigma$,~$\LL[\Sigma]$-formula~$\phi(\bar x, \bar y, \bar z)$, and class of~$\Sigma$-structures~$\CC$, if~$\phi$ has the bipartite order-property on~$\CC$, then there also exists an~$\LL[\Sigma]$-formula~$\psi(x,y,\bar z')$ with singleton variables~$x$ and~$y$ that has the bipartite order-property on~$\CC$.
\end{lemma}

\begin{proof}
    Let~$R_\phi(\bar x, \bar y, \bar z)$ be a new relation symbol, let~$\Sigma^+ := \Sigma \cup \{ R_\phi \}$, and let~$\CC^+$ be the~$\Sigma^+$-expansion of~$\CC$ where we interpret~$R_\phi(\bar x, \bar y, \bar z)$ as~$\phi(\bar x, \bar y, \bar z)$. This means~$\CC^+$ is a class of~$\Sigma^+$-structures.
    Now the (quantifier-free)~$\FO[\Sigma^+]$-formula~$\phi'(\bar x, \bar y, \bar z) := R_\phi(\bar x, \bar y, \bar z)$ has the bipartite order-property on~$\CC^+$.
    Then~$\phi'$ also has the bipartite order-property on~$\Th(\CC^+)$ by the compactness theorem.
    By \cref{thm:simon} there is an~$\FO[\Sigma^+]$-formula~$\gamma(x,y,\bar z')$ with single free variables~$x$ and~$y$ that has the bipartite order-property on~$\Th(\CC^+)$. 
    By definition there is model~$M$ of~$\Th(\CC^+)$ in which~$\gamma$ has the bipartite~$\infty$-order-property.
    For every~$\ell \in \N$, let~$\alpha_\ell$ be the~$\FO[\Sigma^+]$-sentence expressing that there are no elements~$a_1b_1 \ldots a_\ell b_\ell$ and~$\bar c$ witnessing that~$\gamma(x,y,\bar z')$ has the bipartite~$\ell$-order-property.
    For every~$\ell\in\N$ we have~$M\models \neg \alpha_\ell$ and therefore~$\alpha_\ell \notin \Th(\CC^+)$.
    Then for every~$\ell \in \N$ there must be a structure~$G \in \CC^+$ on which~$\gamma$ has the bipartite~$\ell$-order-property.
    Hence,~$\gamma$ has the bipartite order-property on~$\CC^+$.
    Replacing in~$\gamma(x,y, \bar z')$ each occurrence of~$R_\phi$ with~$\phi$ yields the desired~$\LL[\Sigma]$-formula~$\gamma(x,y, \bar z')$ that has the bipartite order-property on~$\CC$.
\end{proof}

We are now ready to prove \cref{lem:mstable-transductions}
\begin{proof}[Proof of \cref{lem:mstable-transductions}]

    Assume first that~$\CC$~$\LL$-transduces the class of all half-graphs.
    Then there exists~$k\in \N$ and an~$\LL[\Sigma^{(k)}]$-formula~$\phi(x,y)$ such that for every~$\ell \in \N$ there is~$G^+ \in \CC^{(k)}$ with elements~$a_1,\ldots,a_\ell \in V(G^+)$ and~$b_1, \ldots,b_\ell \in V(G^+)$ such that for all~$i,j \in [\ell]$ we have~$G^+ \models \phi(a_i,b_j)$ if and only if~$i \leq j$.
    The formula~$\psi(x_1 x_2, y_1 y_2) := \phi(x_1,y_2)$ has the~$\ell$-order-property on~$G^+$ witnessed by the tuples~$\bar a_i := a_i b_i$.
    Hence,~$\psi$ has the order-property in~$\CC^{(k)}$ and~$\CC$ is not monadically~$\LL$-stable.

    Assume now~$\CC$ is not monadically~$\LL$-stable.
    Then there exist~$k \in \N$ and an~$\LL[\Sigma^{(k)}]$-formula~$\phi(\bar x, \bar y)$ that has the (bipartite) order-property on~$\CC^{(k)}$.
    By \cref{lem:singletons}, there exists a formula~$\psi(x,y,\bar z)$ with singleton variables~$x$ and~$y$ that has the bipartite order-property on~$\CC^{(k)}$.
    Constructing a transduction from~$\CC^{(k)}$ to the class of all half-graphs is now standard.
    As~$\CC^{(k)}$ is a coloring of~$\CC$, also~$\CC$ transduces the class of all half-graphs.
\end{proof}

In \cite{simon2021note}, Simon also proves an analog of \cref{thm:simon}, showing that also the \emph{independence-property} (cf.\ \cref{sec:mstable}) is always witnessed by formulas~$\psi(x,y,\bar z)$ with two singleton free variables and parameters.
Following the proofs above, we obtain the following lemma.

\lemMdepTransductions*

\end{document}